%% file: main.tex
\documentclass[twocolumn]{openjournal}

\usepackage[colorlinks=true, linkcolor=blue, citecolor=blue, urlcolor=blue]{hyperref}
\usepackage{amsmath}
\usepackage{natbib}
\usepackage{booktabs}
\usepackage{graphicx}
\usepackage{threeparttable}
\usepackage{orcidlink}

\newcommand{\temp}[2]{$T_{e,[\text{#1 #2}]}$}

\newcommand{\te}{$T_{e}$ }

\newcommand{\oiii}{[O\,\textsc{iii}]}
\newcommand{\nii}{[N\,\textsc{ii}]}
\newcommand{\sii}{[S\,\textsc{ii}]}

\newcommand{\siii}{[S\,\textsc{iii}]}
\newcommand{\oii}{[O\,\textsc{ii}]}

\newcommand{\hii}{H\,\textsc{ii}}
\newcommand{\ha}{H$\alpha$}
\newcommand{\hb}{H$\beta$}
\newcommand{\niiauroral}{[N\,\textsc{ii}]$\lambda$5755}

\defcitealias{RosalesOrtega2026}{RO26}
\defcitealias{Dopita2016}{D16}
\defcitealias{Pilyugin2016}{PG16}
\defcitealias{Kewley2002}{KD02}
\defcitealias{Kobulunicky2004}{KK04}
\defcitealias{P05}{PT05}
\defcitealias{Delgado2023a}{MD23}

\newcommand{\revone}{}
\newcommand{\revtwo}{}

\begin{document}

\title{Exploring the synergies of [\ion{O}{2}]3727 with MUSE spectroscopy in PHANGS \hii\ regions}
\shorttitle{\hii\ regions with SITELLE \oii3727 and MUSE spectroscopy}
\include{authors.tex}

\begin{abstract}
Spatially resolved maps of gas-phase metallicity provide key constraints on the chemical enrichment and mixing processes that drive galaxy evolution, but measurements based only on strong lines remain highly uncertain and dependent on \revtwo{emission line coverage}. In this work, we present a joint analysis of SITELLE observations, covering the \oii$\lambda\lambda$3726,3729 doublet, with PHANGS-MUSE spectroscopy (covering 4800--9300\AA; \revtwo{including}  H$\beta$, \oiii$\lambda$4959,5007, \nii$\lambda$6584, H$\alpha$, \sii$\lambda\lambda$6716,6731, \siii$\lambda$9069) within five nearby spiral galaxies. By combining these, we construct a homogeneous catalog of emission-line fluxes for 604 ionized nebulae, 556 of which are classified as \hii\ regions. This enables a comparison of eight widely used strong line metallicity calibrations, five new strong line calibrations, and an investigation of ionization parameter diagnostics. We recover known systematic offsets among calibrations, but also find that many exhibit very low scatter ($<$0.03--0.04~dex) in radial metallicity gradients. We find that \siii/\sii\ \revtwo{exhibits minimal secondary dependence on metallicity or extinction, thus it may be} a more robust tracer of ionization parameter than \oiii/\oii. No significant outliers are identified in O/H or N/O within the sampled regions, \revtwo{indicating internally consistent abundance trends across the inner disks probed by our data.} \revtwo{We provide a publicly available catalog of all measured emission-line fluxes, designed to support future investigations, including temperature modeling and strong-line abundance calibrations.} 
\end{abstract}


\maketitle

\section{Introduction}
\label{sec:intro}

Heavy elements produced by stellar nucleosynthesis accumulate over cosmic time and are redistributed by spatially inhomogeneous star formation and feedback, producing pronounced variations in chemical abundances within galaxies \citep{Matteucci2012, Maiolino2019A&ARv..27....3M}. The \revone{radial and azimuthal trends in} chemical abundances across the disks of galaxies are constrained by measuring the current gas-phase oxygen abundance (metallicity) within \hii\ regions \citep{Kennicutt1996ApJ...456..504K, 2014AJ....147..131P, Kreckel2019, 2020MNRAS.492.4149S}.  In addition to tracing stellar feedback and chemical enrichment from massive stars into the interstellar medium (ISM), metals play a fundamental role in regulating ISM cooling and shaping the physical conditions of the local ISM (e.g., gas-to-dust ratio, HI-to-H$_2$ transition, CO-to-H$_2$ conversion factor).  The predominantly negative radial trends in metallicity have been well established across large samples of galaxies \citep{Zaritsky1994ApJ...420...87Z, Moustakas2010, Sanchez2014A&A...563A..49S, Kaplan2016MNRAS.462.1642K,Belfiore2017MNRAS.469..151B, Poetrodjojo2018MNRAS.479.5235P}, tracing the inside-out growth of galaxy disks \citep{Boissier1999MNRAS.307..857B}, while mapping azimuthal variations has remained challenging because it requires resolving small-scale abundance structure across the disk with sufficient sensitivity and spatial coverage \citep{Kreckel2019, 2020MNRAS.492.4149S, 2022MNRAS.509.1303W, Bresolin2025MNRAS.539..755B}. 

With the introduction of wide-field optical integral field unit (IFU) spectrographs, it has become feasible to map metallicities for hundreds of \hii\ regions across individual galaxy disks \citep{2019MNRAS.484.5009E, 2020MNRAS.494.1622E}.
\revone{With this goal in mind, } the Physics at High Angular resolution in Nearby GalaxieS (PHANGS) collaboration has completed a large observing program employing the Very Large Telescope/Multi Unit Spectroscopic Explorer (VLT/MUSE; \citealt{ESOMUSE}) to mosaic the disks of 19 nearby ($D<$19 Mpc), low-inclination spiral galaxies \citep{Emsellem2022}.
The PHANGS sample constitutes one of the largest, homogeneous samples of \hii\ regions in external galaxies, a fundamental resource for future studies.  Systematic azimuthal variations are observed in half of the sample \citep{Kreckel2019, 2022MNRAS.509.1303W}.
Interestingly, these maps also reveal low ($\sigma_{O/H}\sim$0.04 dex) scatter in oxygen abundances, indicative of efficient mixing, and correlations with local conditions indicating recent star formation may have locally enriched the material \citep{Kreckel2020MNRAS.499..193K}. 
\revone{These have been measured using indirect \revtwo{`strong line'} methods, as a direct determination of the metallicity requires knowledge of the physical gas conditions, particularly the electron temperature and density. }
However, there is a long-standing debate about how to calculate metallicity when the gas electron density and temperature cannot be directly measured \citep{Kewley2019, Maiolino2019A&ARv..27....3M}. Using only the strongest emission lines, offsets of up to 0.5~dex are found between calibrations \citep{Kewley2008, Croxall2013, Blanc2015ApJ...798...99B}.

\revone{Different metallicity calibrations are known to result in pronounced differences in the resulting metallicity measurements \citep{Kewley2008, 2021MNRAS.507.2468S, Groves2023}, but pragmatically specific calibrations are commonly selected based on the available emission lines.} 
The red wavelength coverage of MUSE (4800--9300 \AA) limits the number of strong emission lines available for determining the metallicity in nearby galaxies (i.e.~H$\beta$, \oiii$\lambda$4959,5007, \nii$\lambda$6584, H$\alpha$, \sii$\lambda\lambda$6716,6731, \siii$\lambda$9069), and misses in particular the blue line \oii$\lambda\lambda$3726,3729 (hereafter \oii) that would provide insights into a number of ISM physical conditions. \revtwo{While MUSE does cover the \oii$\lambda\lambda$7320,7330 doublet, it is significantly fainter and only detected in $\sim$1\% of \hii\ regions \citep{Brazzini2024A&A...691A.173B}.} 
With the addition of \oii$\lambda\lambda$3726,3729, it is possible to measure the diagnostic line ratio $R_{23}\equiv$(\oii$\lambda$3727+3729 + \oiii$\lambda$4959+5007)/H$\beta$, which is necessary for some of the most common metallicity calibrations \citep{Kewley2002, Pettini2004, Kobulunicky2004, P05, Pilyugin2016}.  A comparison of \oiii/\oii\ also provides insights into the ionization parameter of a given \hii\ region, and facilitates a direct measurement of the electron temperature of O$^+$ (requiring both \oii$\lambda$7320,7330 and \oii$\lambda\lambda$3726,3729).    
Finally, the \oii\ line is crucial for determining the N/O ratio, which is often assumed to be constant \revtwo{at low gas-phase metallicity and slightly increasing at higher metallicites, but} might change systematically depending on the star formation, accretion, and gas ejection history of galaxies \citep{PerezMontero2009MNRAS.398..949P, Belfiore2017MNRAS.469..151B, Berg2020ApJ...893...96B, Stiavelli2025ApJ...981..136S}. 

To obtain measurements of the crucial \oii\ line, we analyze observations from the Spectromètre Imageur à Transformée de Fourier pour l'Étude en Long et en Large de raies d'Emission (SITELLE; \citealt{Grandmont2012SPIE.8446E..0UG}), an instrument on the Canada-France-Hawaii Telescope (CFHT). It provides integral field unit (IFU) spectroscopic capabilities in the visible (350 to 900 nm) over an 11 by 11 arcminutes field of view and, by selecting the appropriate filter (SN1), it is possible to obtain IFU maps of \oii$\lambda$3727,3729. 
Of the 19 PHANGS-MUSE galaxies, five have been observed with SITELLE, three of these (NGC\,628, NGC\,3351, NGC\,3627) as part of the SIGNALS large program \citep{Rousseau-Nepton2019MNRAS.489.5530R}. 
\revtwo{In this paper, we combine our measurements of the \oii\ line from SITELLE with the measurements of the redder lines from PHANGS-MUSE data to refine gas-phase metallicity measurements across five nearby galaxies using the empirical calibrations inaccessible with the MUSE or SITELLE data only. }

In Section \ref{sec:sample} we present an overview of the galaxy sample, and the catalog of ionized nebulae that is the basis for our analysis of the \oii\ line emission. In Section \ref{sec:data}, we present our new observations. In Section \ref{sec:results} we show results that leverage the \oii\ detections, and in Section \ref{sec:discussion} we discuss these in the context of the strong line calibrations available. Finally, we conclude and summarize our findings in Section \ref{sec:conclusions}.  
Throughout this paper, we use the \revtwo{notation for emission lines and line ratio diagnostics in Table \ref{tbl:line_ratio_definitions}}.

\begin{table}
\caption{Emission line and line ratio notation.}
\centering
\footnotesize
\begin{tabular}{cc}
\toprule
Quantity & Definition \\
\midrule
\oii & \oii$\lambda\lambda$3726,3729 \\
\oiii & \oiii$\lambda$5007 \\
\nii & \nii$\lambda$6583 \\
\sii & \sii$\lambda$6717 + \sii$\lambda$6731 \\
\siii & \siii$\lambda$9069 \\
\midrule
R2 & $\log$(\oii/\hb) \\
R3 & $\log$(\oiii/\hb) \\
R23 & $\log$((\oii+\oiii$\lambda\lambda$4959, 5007)/\hb) \\
$\hat{R}$ & 0.47 $\times$ R2 + 0.88 $\times$ R3 \\
O32 & $\log$(\oiii/\oii) \\
N2O2 & $\log$(\nii/\oii) \\
N2S2 & $\log$(\nii/\sii) \\
N2 & $\log$(\nii/\ha) \\
N2S2Ha & N2S2
+ 0.264 N2 \\
\bottomrule
\end{tabular}
\begin{tablenotes}
\centering
\small
\item \textbf{Notes.} All logarithms are base 10.
\end{tablenotes}
\label{tbl:line_ratio_definitions}
\end{table}

\section{Galaxy Sample and Nebular Catalog}
\label{sec:sample}

This study focuses on the only five galaxies from the PHANGS-MUSE sample \citep{Emsellem2022} that have SITELLE observations available covering the \oii$\lambda$3727 emission line. These five galaxies (Table \ref{tbl:sample_observations}) are all nearby (D$<$20~Mpc) star-forming spiral galaxies. With typical ground-based $\sim$1\arcsec\ seeing, this corresponds to a physical resolution $<$100~pc, sufficient to isolate individual \hii\ regions from their neighbors and their surroundings.

\begin{table*}[!t]
\caption{General properties and SITELLE observing parameters for galaxies in our sample.}
\centering
\footnotesize
\begin{tabular}{cccccccccc}
\toprule
Galaxy &  Distance$^{1}$ & $v_{\rm sys}$$^{2}$ & log$_{10} M_{\star}$$^{3}$ & Log(SFR)$^{3}$ & Observing & Exposure & Resolution & PSF FWHM & N tot (\hii)\\
Name &  [Mpc] & [km s$^{-1}$] & [$M_{\odot}$] & [$M_{\odot} yr^{-1}$] & Date & Time [s] & [$\lambda / \Delta \lambda$] & [arcsec] & \\
\midrule
NGC~628 &  9.84 & 651 & 10.34 & 0.24 & 2016\slash01\slash13 & 7665 & $\sim$600 & 1.20 & 235 (207)\\
NGC~2835 &  12.2 & 867 & 10.00 & 0.09 & 2020\slash02\slash25 & 6690 & $\sim$1250 & 1.48 & 129 (115)\\
NGC~3351 & 10.0 & 775 & 10.36 & 0.12 & 2019\slash04\slash09 & 10089 & $\sim$950 & 1.34 & 29 (18)\\
NGC~3627 &  11.3 & 715 & 10.83 & 0.58 & 2021\slash12\slash16 & 7552 & $\sim$650 & 1.41 & 223 (205)\\
NGC~4535 & 15.8 & 1954 & 10.53 & 0.33 & 2020\slash02\slash21 & 7992 & $\sim$1250 & 1.40 & 9 (9)\\
\bottomrule
\end{tabular}
\begin{tablenotes}
\centering
\small
\item \textbf{Notes.} $^{1}$From \cite{Anand2021}.  $^{2}$From \cite{Makarov2014}.  $^{3}$From \cite{Leroy2021}.
\end{tablenotes}
\label{tbl:sample_observations}
\end{table*}

Our measurements of \oii\ are all based on the regions defined in the PHANGS-MUSE nebular catalog
constructed by \cite{Groves2023}. 
Using optical integral field spectroscopy from the MUSE on the VLT, those authors used the H$\alpha$ line emission morphology to identify 8,847 distinct ionized nebulae across these five galaxies. For each of these, an integrated spectrum is extracted from the reduced MUSE data cube \citep{2020A&A...641A..28W} and fit using the PHANGS-MUSE Data Analysis Pipeline \citep[as detailed in][]{Emsellem2022}. \revone{The \cite{Groves2023} catalog contains measurements of} Gaussian-fit line fluxes corresponding to the brightest emission lines (\hb, \oiii$\lambda$5007, \ha, \nii$\lambda$6583, \sii$\lambda\lambda$6716,6731, \siii$\lambda$9069).  
All reported line fluxes \revtwo{from the nebular catalog} are by default corrected for Milky Way foreground extinction, assuming the E(B - V) values provided by \cite{Schlafy2011} and an \cite{ODonnnell1994} extinction law. Line fluxes that are corrected for extinction internal to the galaxy are also calculated by comparing the observed Balmer decrement (\ha/\hb) to the theoretical value assuming an intrinsic Balmer ratio of \ha/\hb\ = 2.86, an \cite{ODonnnell1994} extinction law, and R$_V$=3.1 for all regions. \revtwo{After correcting for reddening, \cite{Groves2023} uses} diagnostic line ratios 
(e.g. BPT; \citealt{Baldwin1981, Veilleux_1987}) to classify the photoionized subset of nebulae as `\hii\ regions'. Regions that fall below the \cite{Kauffmann2003} diagnostic curve in the \oiii/\hb\ versus \nii/\ha\  diagram and below the \cite{Kewley2006} diagnostic curve in the \oiii/\hb\ versus \sii/\ha\  are flagged as \hii\ regions. \revtwo{Finally, in this work, we require a S/N$>$5 of all strong emission lines used in BPT diagnostics.} This results in a parent sample of  6,300 as \hii\ regions. 

\begin{figure}
\centering
  \includegraphics[width=0.475\textwidth]{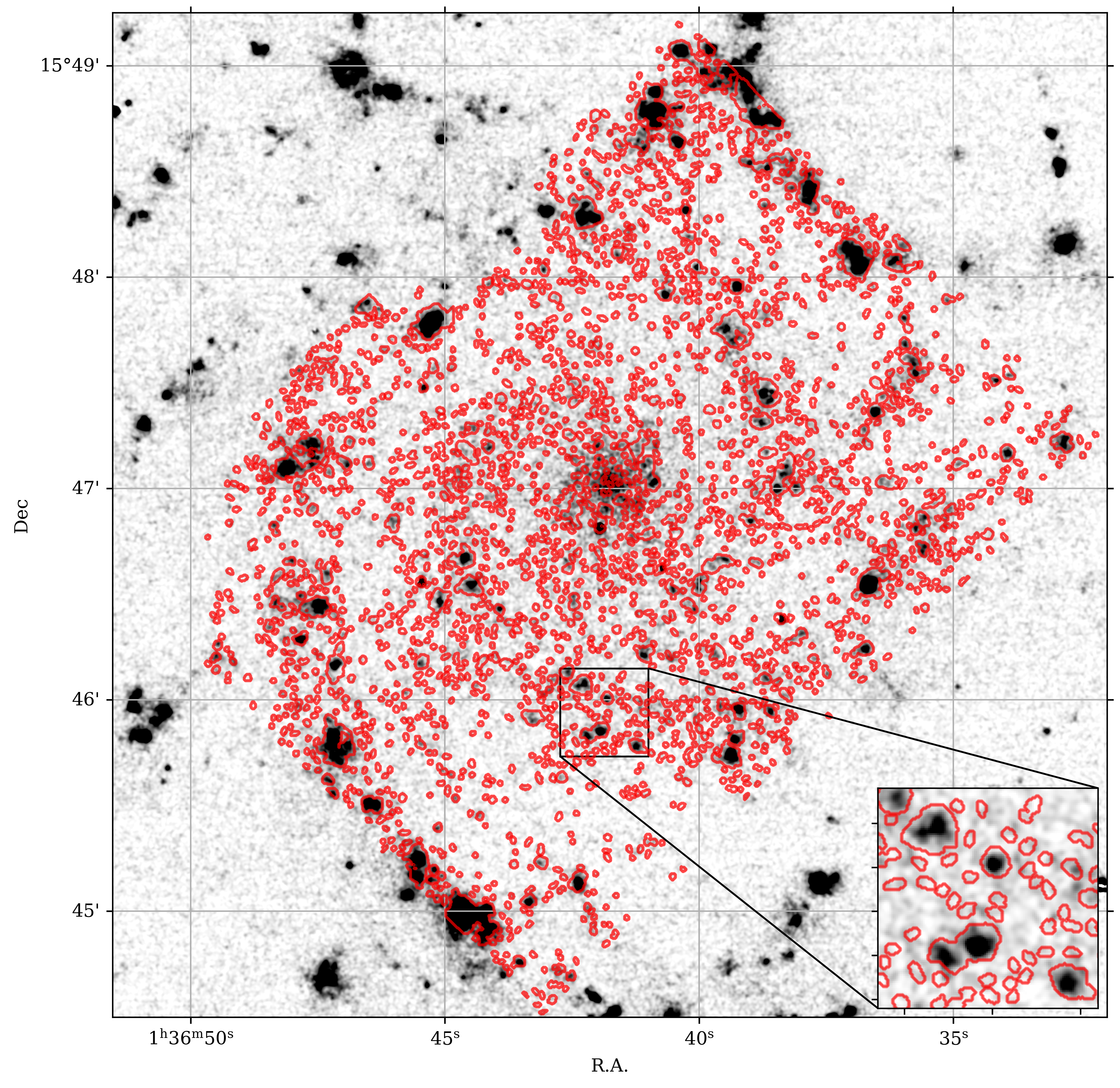}
  \caption{The NGC~628 \hii\ region mask obtained from the Nebular Catalog \citep{Groves2023} is overlaid in red on the \oii\ flux map. An \oii\ velocity map was created using SNR $>$ 5 detections of \oii. This velocity map was then extrapolated to the same dimensions as the MUSE image using a nearest-neighbor algorithm. Using the velocity map, we found the redshifted \oii\ and extracted the integrated flux density within $\pm$2 \AA\ of the feature.}
  \label{fig:HIImask}
\end{figure}

\cite{Brazzini2024A&A...691A.173B} carried out a careful emission line fitting of the faint \niiauroral\ auroral line in these MUSE spectra, resulting in robust S/N $>$3 detections in 91 of these \hii\ regions. As in \cite{Kreckel2025A&A...703A..42K}, we use the \texttt{pyneb} package \citep{Luridiana2015} to combine the extinction-corrected \niiauroral\ and \nii$\lambda$6583 lines and compute the electron temperature for N$^{+}$ based on the \nii\ lines (\temp{N}{II}), using the \sii\ line ratio to derive the electron density. \revtwo{The \temp{N}{II} measurements from \cite{Brazzini2024A&A...691A.173B} are then used to derive metallicities using the relation from \cite{Delgado2023a}; these metallicities are analyzed in Section~\ref{subsec:strong_line_results}.}

\section{SITELLE Data}
\label{sec:data}

\subsection{Observations and Data Reduction}

Five nearby galaxies were observed by SITELLE \citep{Grandmont2012SPIE.8446E..0UG} at the Canada-France-Hawaii Telescope (CFHT); two of these galaxies (NGC~2835, and NGC~4535) were observed by the PHANGS collaboration (proposal ID 20AF06; PI: Hughes) and three (NGC~628, NGC~3351, and NGC~3627) by the Star formation, Ionized Gas, and Nebular Abundances Legacy Survey (SIGNALS, proposal ID 20BP41; PI:  Rousseau-Nepton). The properties of the galaxies in these observations are seen in Table \ref{tbl:sample_observations}. 
The SITELLE observations used in this study make use of SITELLE's SN1 filter that covers the 3650 [\AA] $\sim$ 3850 [\AA] range; Table \ref{tbl:sample_observations} provides an overview of the observed data.

Each SITELLE data cube was processed by the Outil de Réduction Binoculaire pour SITELLE (ORBS) reduction package \citep{ORBS_ORCS}. The calibration and reduction process described below follows that of \cite{Rousseau-Nepton2018}. The flux calibration for imaging of each galaxy was performed using observations of a known spectro-photometric standard star (LDS749B for NGC~2835, NGC~3351 and NGC~4535; GD 71 for NGC~628; Hz 21 for NGC~3627). \revtwo{A white light calibration cube was obtained and used to characterize high-order phase variations across each field, which were then corrected during the ORBS data reduction process.}

\subsection{Alignment}\label{subsec:alignmet}

We aligned the SITELLE cubes to the MUSE astrometry by deriving an average spatial offset by comparing bright sources in both datasets. These sources were initially identified with \texttt{DAOStarFinder} from \texttt{photutils} \citep{photutils} in the SITELLE deep-frame images\revtwo{, constructed by summing the flux density over all wavelength channels, and in a blue-weighted MUSE image constructed using the portion of the Johnson B transmission curve overlapping the MUSE spectral range \citep{Emsellem2022}. This MUSE image was used only for source detection and centroiding during the astrometric alignment, as the MUSE spectral range does not cover the full Johnson B bandpass. The image should not be interpreted as a true synthetic Johnson B image and was not used for photometric calibration.} Moreover, to reduce centroid uncertainties introduced by the wavelength difference between the two images, we manually selected only bright, isolated sources with point-like morphology.

An effective Point Spread Function (ePSF) was created for both the SITELLE and MUSE images in each galaxy using the \texttt{EPSFBuilder} method from \texttt{photutils}. In NGC~628 we used 3 sources to create this ePSF; 9 in NGC~2835; 14 in NGC~3351; 11 in NGC~3627; 3 in NGC~4535. The FWHM of the ePSFs are reported in Table \ref{tbl:sample_observations}. Each bright source was fitted using the resultant ePSF, which resulted in new centroids of each bright source. The difference in R.A. and Dec. between bright sources in each image was found, and the median difference in degrees was applied to the astrometry in the SITELLE data cubes. 

\subsection{Spectra Extraction from SITELLE Cubes}\label{subsec:spec_extraction}

A sky subtraction, similar to the one done in \cite{Rousseau-Nepton2018}, was performed for each SITELLE data cube in our sample. An annulus is defined far outside each galaxy in order to avoid an overlap with flux emitted from the galaxy. We extract a spectrum from each pixel contained within each annulus and take the median flux density for all wavelength channels. We tried varying the size of the annulus and found that across the five galaxies, the background spectrum had an average flux value of $32.31\pm0.05$ $[10^{-20}\ \text{erg} / \text{cm}^2 / \text{s} / \text{\AA}]$. The small standard deviation for all background spectra indicates that this is an unbiased method for measuring the sky background in the cube. The median sky background spectrum is then subtracted from every pixel in the respective SITELLE data cubes.

The aligned and sky background subtracted SITELLE cubes were then reprojected into the same pixel dimensions as the MUSE images using the \texttt{mProject} method from \texttt{MontagePy}. The \hii\ region masks from the Nebular Catalog can then be applied, which were created using H$\alpha$ emission and the Python package \texttt{HIIphot} \citep{Thilker2000}. Figure \ref{fig:HIImask} shows the \hii\ region boundaries overlaid on the \oii\ emission line map from the NGC~628 SITELLE data cube. We then produced integrated SITELLE spectra for each \hii\ region using the MUSE nebular catalog mask.

\subsection{[OII] \texorpdfstring{$\lambda$3727}{lambda3727} fitting and measurement}
\label{subsec:oii3727}

The background-subtracted \hii\ region spectra were passed into the \texttt{ORBS} \citep{ORBS_ORCS} \texttt{fit\_lines\_in\_spectrum} method to fit the unresolved \oii\ doublet. The \oii\ feature is modeled using a single `sincgauss' shape, which is the convolution of the Gaussian and sinc functional forms. For a detailed explanation of the sincgauss convolution refer to Sections 2 and 3 of \cite{Martin2016}. The spectral resolution of the SITELLE observations near the \oii\ doublet is $\Delta\lambda > 3$ \AA, so the \oii\ doublet is blended in all five galaxies. Early in our analysis, we attempted to measure the two components separately; however, because the doublet does not exhibit spectrally resolved peaks, fitting it as a single feature provides a more reliable measurement of \oii$\lambda$3727. 

\begin{figure*}
\centering
  \includegraphics[width=\textwidth]{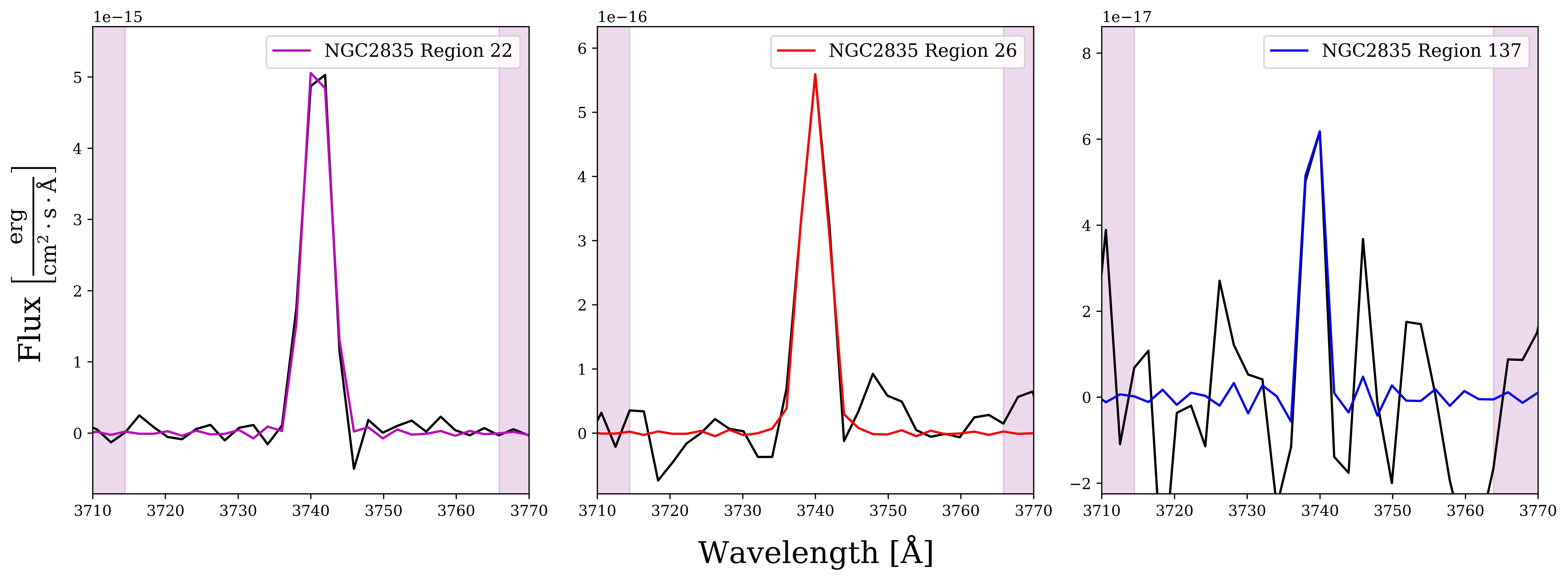}
  \caption{Three examples of spectral fits of \oii$\lambda$3727 made with a `sincgauss' template using the \texttt{fit\_lines\_in\_spectrum} method from the ORCS package \citep{ORBS_ORCS}. Each fit shown is from NGC~2835 with the actual observed spectrum in black, and fits for Region 22 in purple (SNR $\sim$50), Region 26 in red (SNR $\sim$12) and Region 137 in blue (SNR $\sim$3.5). \revone{Note that the wiggles in the model spectrum are expected due to the `sincgauss' line shape.} A portion of the spectral regions where the continuum levels are estimated is shaded in purple (see Section \ref{subsec:oii3727} for details).}
  \label{fig:OII3727}
\end{figure*}

\revtwo{For each \oii\ line, we estimate the local continuum by fitting a linear polynomial to the spectral windows [$\lambda - 60$, $\lambda - 25$] and [$\lambda + 25$, $\lambda + 95$]. This fitted continuum is then subtracted from the entire SN1 filter range (3650--3850~\AA).} We allow the spectral position, velocity dispersion and the amplitude to be free parameters in the \texttt{ORBS} fitting algorithm. We use the \nii $\lambda$6584 velocity from the MUSE spectrum from the same region as an initial guess for the ORBS fitting method. Examples of \oii\ fits are shown in Figure \ref{fig:OII3727}. 

\oii\ fluxes are obtained as an output from the \texttt{fit\_lines\_in\_spectrum} method from ORBS \citep{Martin2016}. We determine uncertainties using an MC method. A Gaussian noise distribution is created with a mean of zero and a standard deviation equal to that of the surrounding continuum. \revtwo{For each wavelength bin, a random draw from this distribution is added to the corresponding flux value.} Our fitting process is carried out 1,000 times with added Gaussian noise, and the standard deviation in the obtained \oii\ fluxes is taken to be the 1 $\sigma$ uncertainty. 

Emission-line flux measurements are considered robust if \revtwo{S/N $>$ 5}. The signal is defined as the fitted amplitude of the sincgauss function, and the noise is taken to be the standard deviation of the background continuum. We correct our \oii\ fluxes for reddening using E(B$-$V) Milky Way foreground extinction values taken from \cite{Schlafy2011} and for local extinction using the E(B$-$V) values for each \hii\ region calculated from the MUSE Balmer decrement obtained from the Nebular Catalog. For both corrections, we use the extinction law from \cite{ODonnnell1994}. 

\subsection{Wavelength Calibration}\label{subsec:wavecal}

The \oii\ line is used to calibrate the wavelength axis of each SITELLE pointing since there are no strong sky lines in the SN1 filter. We assume the center wavelength of the blended feature to be $\sim$3728.02 \AA. This is found by taking the difference of the two lines, dividing the difference by the line intensity ratio in the low-density theoretical limit 1.4 = \oii $\lambda\lambda$ 3729/3726, and adding this value to the blue \oii\ feature at 3726.032 \AA\ \citep{NIST_elines}. An \oii\ $\lambda$3727 radial velocity $v_{\rm OII}$ for each S/N $>$ 5 fit is calculated prior to wavelength calibration using the blended rest frame wavelength. A comparison for each \hii\ region is done between $v_{\rm OII}$ and the H$\alpha$ velocity $v_{\rm H\alpha}$ as well as the \nii$\lambda$6584 velocity $v_{\rm NII}$ from the Nebular Catalog. We take the median of $v_{\rm OII}$ - $v_{\rm H\alpha}$ and $v_{\rm OII}$ - $v_{\rm NII}$ for each of the five galaxies, and find that these two methods agree to within $\sim$2.4 km s$^{-1}$. Since OII and NII should emit radiation from the same ionization volumes of \hii\ regions, we use the median offset from $v_{\rm OII}$ - $v_{\rm NII}$ to calibrate the wavelength axis in each of our SITELLE data cubes. Each of the five galaxies had a consistent wavelength offset, with a median of $\sim1.75$ \AA; this offset in the wavelength is expected from the SITELLE SN1 filter \citep{Rousseau-Nepton2018}. 

\subsection{Comparison with literature data}

The extinction-corrected \oii\ fluxes were compared with KCWI \oii\ fluxes obtained for three of the galaxies in \cite{RickardsVaught2024ApJ...966..130R}, and with the \hii\ regions in NGC~628 studied using the same SITELLE data by \cite{Rousseau-Nepton2018}. Overall, the fluxes derived in this work show good agreement with both studies, although the scatter relative to \cite{Rousseau-Nepton2018} appears to be driven primarily by differences in the reddening correction and treatment of the diffuse ionized gas. These differences are not expected to introduce significant systematic uncertainty, and full details of these comparisons are provided in Appendix \ref{app:lit_validation}. 

In addition, \cite{RickardsVaught2024ApJ...966..130R} carried out follow-up  observations of small sub-sections of NGC~628, NGC~2835 and NGC~3627 using the Keck Cosmic Web Imager (KCWI) on Keck, to obtain coverage of the blue wavelength range 3650–5550 \AA, including the \oii\ line. 99 of these \hii\ regions overlap with our parent sample.

\subsection{Catalog}

A total of 604 nebulae ($\sim$10\%) are detected with S/N $>$ 5 in \oii, of which 556 are classified as \hii\ regions. Only NGC~628, NGC~2835 and NGC~3627 have a significant ($N>$100) number of \hii\ regions per galaxy, while NGC~3351 and NGC~4535 have fewer than 25 detections each (see Table \ref{tbl:sample_observations}). 
For this sample of nebulae, we release a catalog containing line fluxes and associated errors, along with reddening corrected values, as a supplement to the \cite{Groves2023} nebular catalog. 

\section{Results}
\label{sec:results}

We focus our analysis on the 556 \hii\ regions with \oii\ detections in the following analysis of the strong line metallicities, the ionization parameter, the metallicity gradients, and the N/O abundances. We also briefly comment on the 48 nebulae (8\% of \oii\ detections) that are not classified as \hii\ regions, but are detected in our \oii\ catalog. 

\subsection{Strong Line Metallicities}
\label{subsec:strong_line}

We investigate many of the same strong line calibrations discussed in both 
\cite{Kewley2008} and \cite{Teimoorinia2021}, \revone{with individual calibrations detailed in Appendix \ref{app:strong_line_summary}, and listed in Table \ref{tbl:prescriptions}. }\revtwo{We distinguish between theoretical calibrations of strong-line methods, which rely on photoionization models, and empirical calibrations, which are based on direct metallicity measurements.} Metallicities are calculated for all \hii\ regions with SNR $>$ 3 in the required emission lines. A comparison of all resulting metallicities is shown in Appendix \ref{app:strong_line_summary} (Figure \ref{fig:cornerplot}).

\begin{figure}
\centering
  \includegraphics[width=0.45\textwidth]{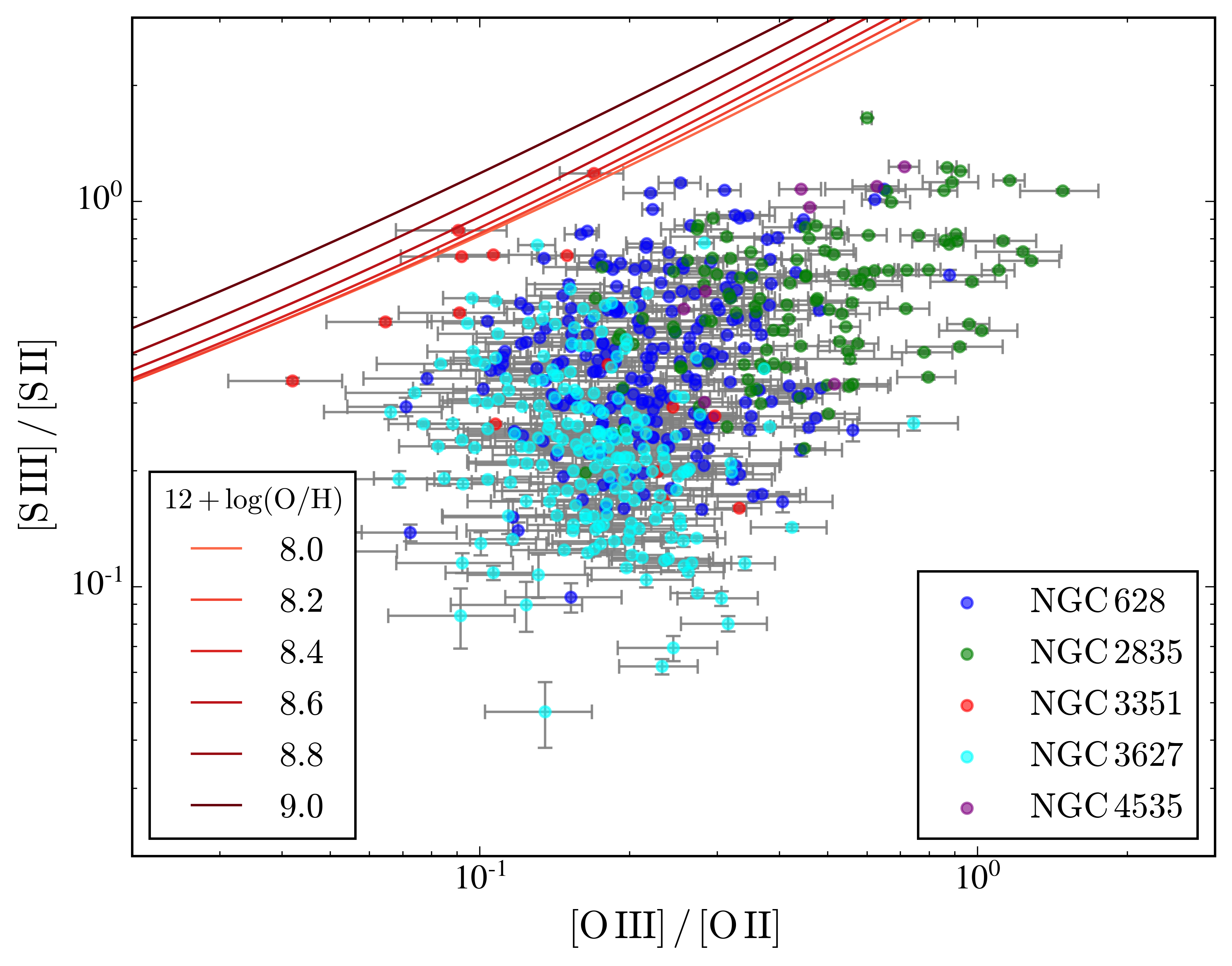}
  \caption{A comparison of the two line ratios, \oiii/\oii\ and \siii/\sii, expected to trace changes in ionization parameter. \hii\ regions in each galaxy are indicated with a separate color (points). \revone{Photoionization models (solid lines; \citealt{Kewley2019}) at a range of metallicities predict a significantly tighter correlation, with the range of metallicities in the model unable to explain the scatter, and are also offset from the observed line ratios.}}
  \label{fig:ion_param}
\end{figure}

\begin{figure*}
\centering
  \includegraphics[width=\textwidth]{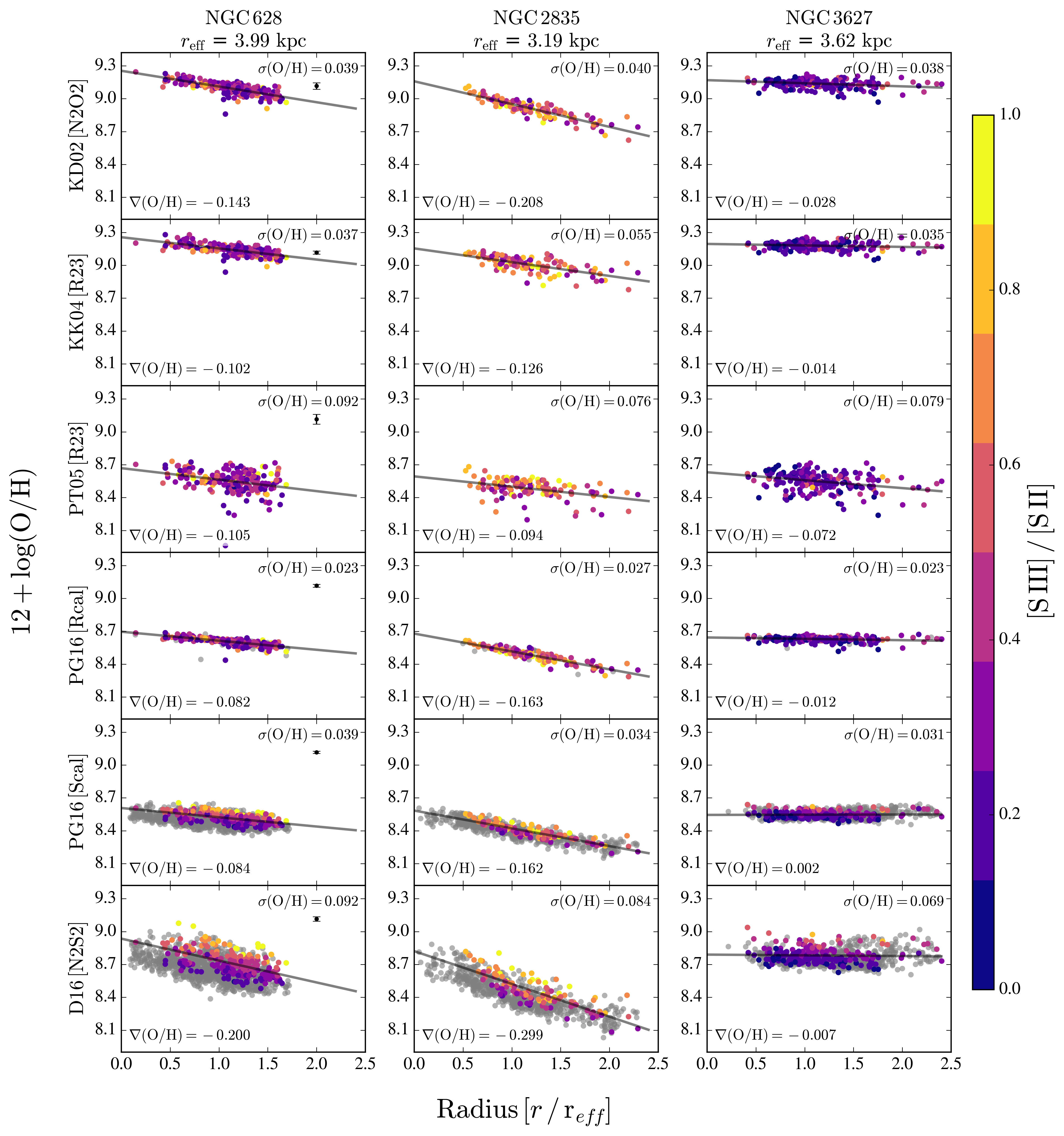}
  \caption{Metallicity gradients for NGC~628, NGC~2835 and NGC~3627 for a sample of the strong line metallicity calibrations. \hii\ regions are colored by their \siii/\sii\ values, and a linear fit is overplotted in black. The standard deviation about that fit ($\sigma$(O/H)) is shown in the top right corner \revtwo{and the slope of the fit is shown in the bottom left corner ($\nabla$(O/H)). R}epresentative error bars shown just below. For calibrations that do not rely on \oii, results from the full Nebular Catalog are also shown in grey. All plots are shown for the same fixed x-axis range in $r_{\rm eff}$ and a fixed y-axis range in 12+log(O/H) spanning 1.5~dex. This ensures all slopes are directly comparable, despite absolute offsets between calibrations. } 
  \label{fig:strong_line_grad}
\end{figure*}

\begin{table*}
\centering
\footnotesize
\caption{Overview of Strong Line Calibrations}
    \begin{tabular}{c|c|c|c}
\toprule
        Abbreviation & Reference & Diagnostic(s) & Method \\
        \hline \hline
        \multicolumn{3}{l}{General Literature Calibrations:} \\
        \hline \hline
        KD02 [N2O2] & \cite{Kewley2002} & N2O2 & theoretical \\
        KK04 [R23] & \cite{Kobulunicky2004} & R23 & theoretical \\
        PT05 [R23] & \cite{P05} & R23 & empirical \\
        PG16 [Rcal] & \cite{Pilyugin2016} & R3, R2, N2 & empirical \\
        PG16 [Scal] & \cite{Pilyugin2016} & R3, S2, N2 & empirical \\
        D16 [N2S2] & \cite{Dopita2016} & N2S2 & theoretical \\
         \hline \hline
         \multicolumn{3}{l}{DESIRED Calibrations:} \\
         \hline \hline
         RO26 [R23] & \cite{RosalesOrtega2026} & R23 & empirical \\
         RO26 [$\hat{R}$] & \cite{RosalesOrtega2026} & R2, R3 & empirical \\
         RO26 [N2O2] & \cite{RosalesOrtega2026} & N2O2 & empirical \\
         RO26 [N2S2] & \cite{RosalesOrtega2026} & N2S2 & empirical \\
         RO26 [N2S2Ha] & \cite{RosalesOrtega2026} & N2S2, N2 & empirical \\
\bottomrule         
    \end{tabular}
    \label{tbl:prescriptions}
\end{table*}

\begin{figure*}
\centering
  \includegraphics[width=\textwidth]{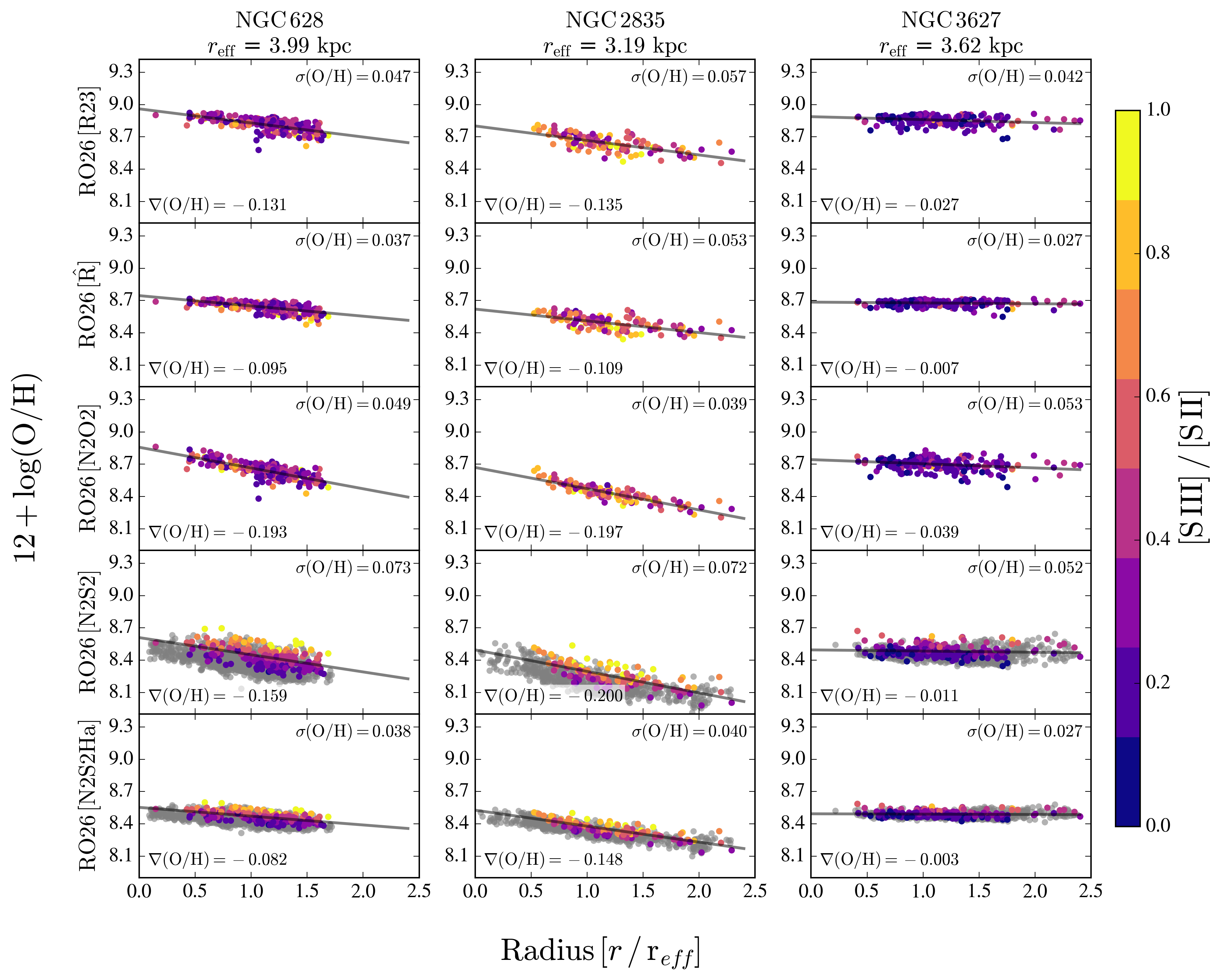}
  \caption{Metallicity gradients for NGC~628, NGC~2835 and NGC~3627 for a sample of the strong line metallicity calibrations provided in \citetalias{RosalesOrtega2026}. \hii\ regions are colored by their \siii/\sii\ values, and a linear fit is overplotted in black. The standard deviation about that fit ($\sigma$(O/H)) is shown in the top right corner \revtwo{and the slope of the fit is shown in the bottom left corner ($\nabla$(O/H)). R}epresentative error bars are shown just below. For calibrations that do not rely on \oii, results from the full Nebular Catalog are also shown in grey. All plots are shown for the same fixed x-axis range in $r_{\rm eff}$ and a y-axis range in 12+log(O/H) spanning 1.5~dex. This ensures all slopes are directly comparable, despite absolute offsets between calibrations.}
  \label{fig:strong_line_grad_DESIRED}
\end{figure*}

We also consider a new set of strong-line calibrations determined using over 2000 high-quality \hii\ region line fluxes from the DESIRED database \citep{Delgado2023b}, as parameterized in  \citet[][hereafter RO26]{RosalesOrtega2026}. As the sample naturally covers a wide range of ionization conditions, over a uniquely wide range of line ratios and metallicities, it has the potential to provide a more uniform approach to strong line metallicity studies. For this work, we consider five of the strong line calibrations, assuming no temperature inhomogeneities ($t^2=0$), and distinguish them by their characteristic line ratios:  
R23, $\hat{R}$, N2O2, N2S2, N2S2Ha.  \revtwo{See Table \ref{tbl:line_ratio_definitions} for notational definitions and Table \ref{tbl:prescriptions} for an overview of each calibration.}

\subsection{Ionization Parameter}

The detection of two nebular lines from different ionization states for oxygen and sulfur allow the ionization parameter tracer to be calculated for each of these elements. The oxygen-based ionization parameter diagnostic 
is defined as 
\oiii/\oii\ $\equiv$ \oiii $\lambda\lambda4959,5007$ / \oii\ $\lambda3727$ and the sulfur ionization parameter tracer is defined as \siii/\sii\ $\equiv$ \siii\ $\lambda\lambda9069,9532$ / \sii\ $\lambda\lambda6717,6731$.  Here we observe only the shorter wavelength \siii$\lambda$9069 line, but adopt a fixed ratio of \siii $\lambda$9532 = $\rm 2.5 \times \siii \lambda$9069, as determined by atomic physics \citep{Osterbrock2006, Tayal2019}. These two tracers for the ionization parameter are plotted against each other in Figure \ref{fig:ion_param}, and \revtwo{we find only a mild correlation between them} (Spearman $\rho$ =  0.496; p$\ll$0.01). \revone{In comparison to photoionization models \citep{Kewley2019}, calculated across a range of metallicities and at fixed pressure\revtwo{, $\log(P/k_{\rm B}) = 5$ in units of ${\rm cm^{-3}\,K}$}, the two ratios show significantly larger scatter than predicted, as well as a 0.5~dex offset that has previously been reported in the literature \citep{Mingozzi2020A&A...636A..42M}. }

\subsection{Metallicity Gradients}
\label{sec:gradients}

In Figure \ref{fig:strong_line_grad} we directly compare the metallicity gradients for the three galaxies with $>$100 \hii\ regions detected (NGC~628, NGC~2835 and NGC~3627), using a selection of the available strong line calibrations (see Table \ref{tbl:prescriptions}). Figure \ref{fig:strong_line_grad_DESIRED} shows the metallicity gradients derived from the DESIRED \citetalias{RosalesOrtega2026} strong line calibrations. \revtwo{All panels are shown on a common x-axis, with the y-axis fixed to span 7.9--9.4 dex.} As a result, it is possible to directly compare the slope and the scatter of each radial trend between calibrations and between galaxies. A simple linear fit to all \oii\ detected \hii\ regions is shown in black, and the scatter with respect to that fit ($\sigma$(O/H)) is listed in the top right of each panel \revtwo{and the slope ($\nabla$(O/H)) in the bottom left of each panel}. All points are colored by their \siii/\sii\ line ratio as a proxy for the ionization parameter. When a calibration does not require the detection of \oii, then the additional \hii\ regions from the nebular catalog are also shown in the background in grey.   

In general, we see qualitatively good agreement between the slopes in all calibrations except those using N2S2 (\citetalias{Dopita2016}~[N2S2] and \citetalias{RosalesOrtega2026}~[N2S2]), which seem to result in a larger dynamic range of values (also seen in Figure \ref{fig:cornerplot}), and produces steeper metallicity gradients in each galaxy as well as a larger value of $\sigma$(O/H). \revone{\citetalias{RosalesOrtega2026}~[N2O2] also shows a somewhat steeper slope. } Across all calibrations, NGC~2835 shows a slightly steeper gradient than NGC~628, and both galaxies show no strong indications of a break or knee in the radial trend out to $\sim$1.5~$r_{\rm eff}$. NGC~3627 exhibits a nearly flat metallicity gradient, with \citetalias{P05}~[R23] producing the steepest slope, $\nabla({\rm O/H}) = -0.072$, and most other calibrations yielding $\nabla({\rm O/H}) \sim 0$. We note that our MUSE coverage limits us to the central bar-dominated portion of the disk. Since bars can drive radial gas flows and enhance mixing within the disk, the measured gradient may be flatter than would be inferred from a larger radial coverage of the galaxy (e.g., \citealt{Zurita2021}).

A secondary correlation with \siii/\sii, where \hii\ regions with higher line ratios show higher metallicity at fixed radius, is apparent in \citetalias{Dopita2016}~[N2S2] and hinted at in \citetalias{Pilyugin2016}~[Scal]. This is suggestive of a missing dependence on ionization parameter that is not well characterized in these calibrations, although we note that the overall scatter in \citetalias{Pilyugin2016}~[Scal] is not larger than what is seen in calibrations where no secondary \siii/\sii\ trend is seen (e.g. \citetalias{Kewley2002}, \citetalias{Kobulunicky2004}). 

\subsection{N/O Abundance Ratios}

\begin{figure}
    \centering
    \includegraphics[width=0.95\linewidth]{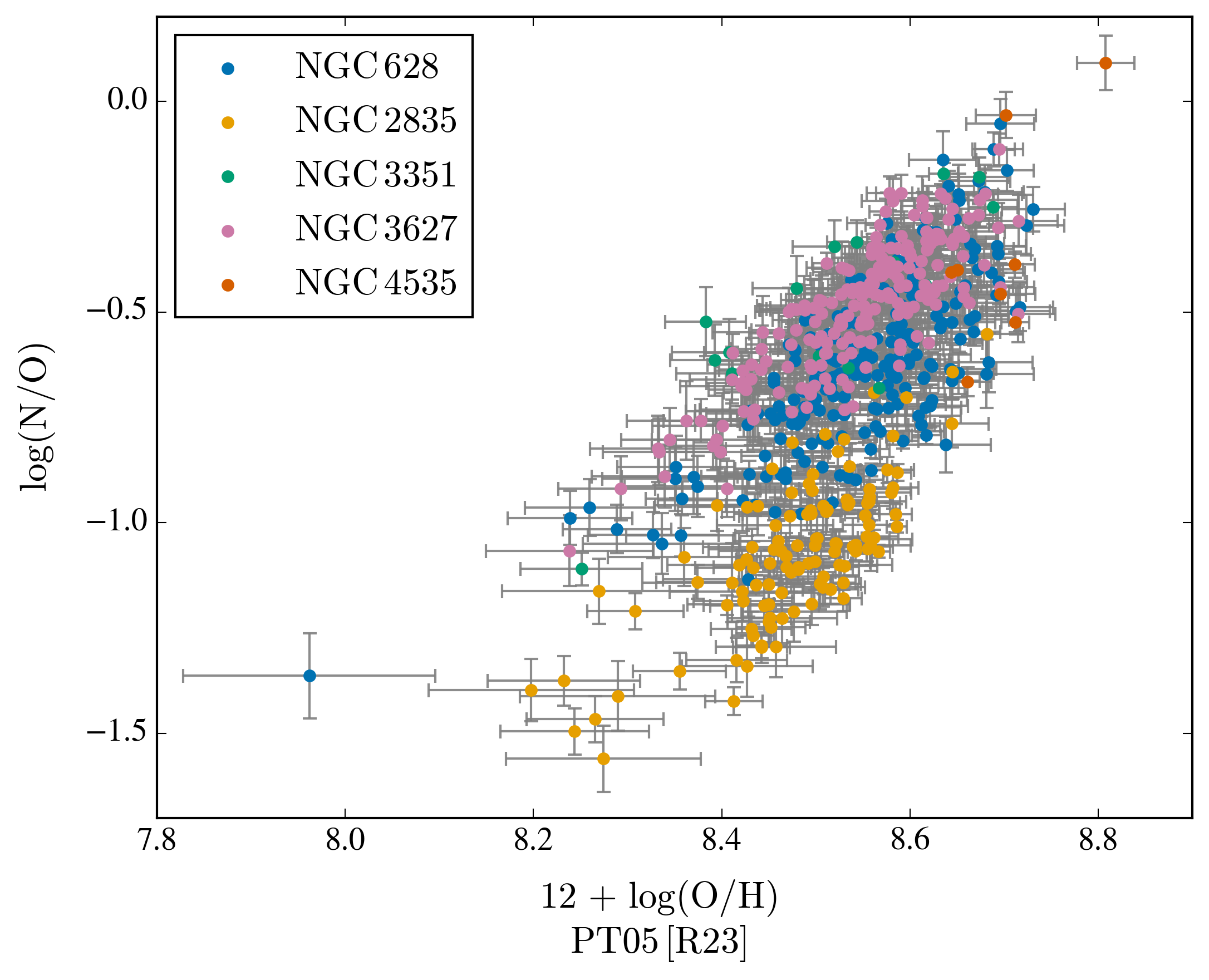}
    \caption{ N/O as a function of 12+log(O/H). Here, we constrain the metallicity using the \citetalias{P05}~[R23] calibration, as it is not reliant on an assumed N/O. Each galaxy is shown with a separate color. }
    \label{fig:NO-OH}
\end{figure}

To explore variations in N/O, we adopt the N/O calculation based on \citetalias{Pilyugin2016}, which relies on \nii/\oii\ and \oii/\hb.  In the high metallicity regime we are probing, N/O is expected to principally trace changes in metallicity because nitrogen has been enriched through secondary production at higher metallicity. However, to compare N/O with O/H, it is important to select an oxygen metallicity calibration that has no built-in assumptions about N/O (common in many theoretical calibrations) or dependence on N2O2 itself \citep{Maiolino2019A&ARv..27....3M}. From the available calibrations, \citetalias{P05}~[R23] is an empirical calibration based principally on R23, and thus the least likely to be biased. In Figure \ref{fig:NO-OH} we see a clear correlation between N/O and O/H using the adopted calibrations. 
Offsets are seen between galaxies, suggestive of different star-formation histories imprinting signatures on the enrichment patterns \citep{Berg2020ApJ...893...96B}. However, more careful treatment of the co-determination of both N/O and O/H would be necessary to ensure these trends are robust \citep{Bresolin2025MNRAS.539..755B}.

\section{Discussion}
\label{sec:discussion}

\subsection{Comparison of strong line methods}
\label{subsec:strong_line_results}

The cross-comparison of eight different strong line metallicity calibrations in Figure \ref{fig:cornerplot} highlights the ongoing challenge in precisely determining absolute metallicity measurements, even for individual \hii\ regions. While more recent calibrations tend to have offsets smaller than $\sim$0.3~dex, well used calibrations in the literature (e.g. \citetalias{Kobulunicky2004}~[R23] and \citetalias{Kewley2002}~[N2O2]) show larger $\sim$0.5~dex offsets. These larger offsets may be driven, at least in part, by the use of older solar oxygen abundance values that are higher than modern estimates; the solar abundances adopted by each calibration are listed in Appendix~\ref{app:strong_line_summary}. However, a more rigorous assessment is required to confirm the origin of these offsets. Nevertheless, as in \cite{Groves2023}, we caution that users of these calibrations should be aware of these differences and only compare with literature results using matched calibrations. 

\begin{figure*}
\centering
\includegraphics[height=4.5cm]{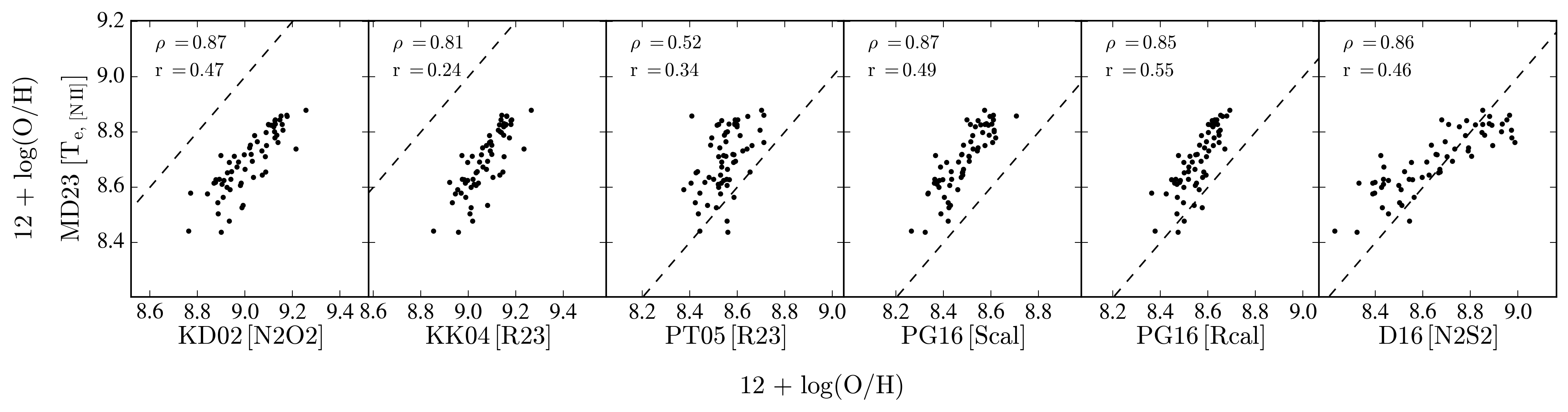}
\includegraphics[height=4.5cm]{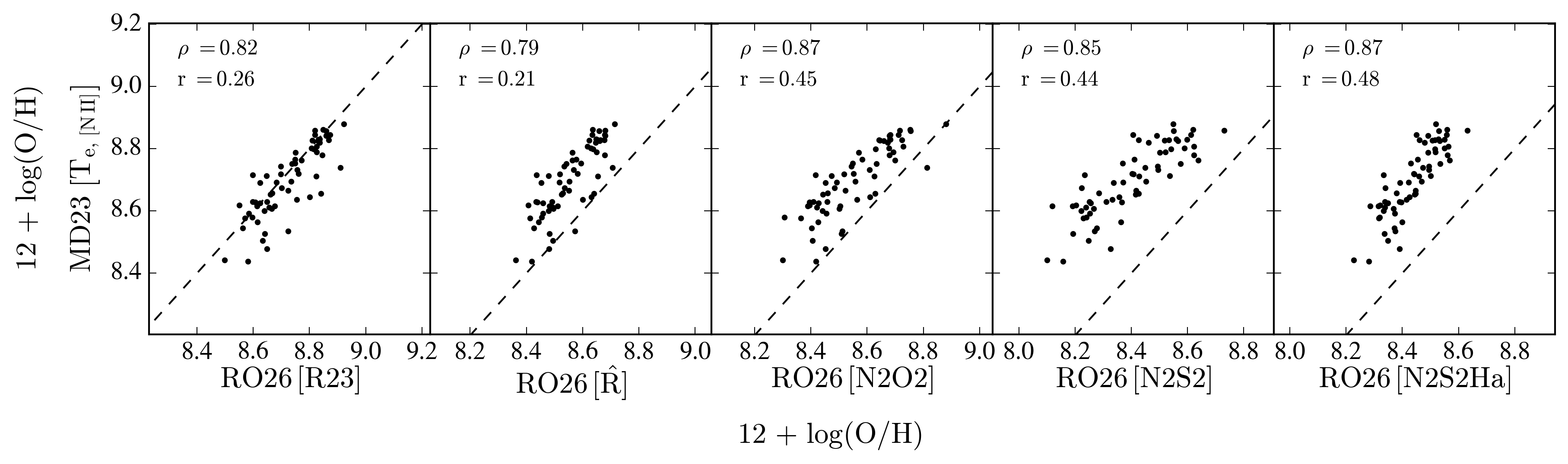}
\caption{Comparison of the \te-based \citetalias{Delgado2023a} calibration with six selected strong-line calibrations (top row) and five selected DESIRED strong-line calibrations (bottom row). Each panel shows the Spearman rank coefficient $\rho$ and Pearson correlation coefficient $r$, which quantify monotonic and linear agreement, respectively; all correlations have $p < 0.01$. Grey dashed lines indicate one-to-one agreement, highlighting absolute offsets between calibrations.}
\label{fig:strong_line_comp}
\end{figure*}

Recent work by \cite{Delgado2023a}, hereafter MD23~[\temp{N}{II}], directly derived metallicities using deep spectra from extragalactic and Galactic \hii\ regions. They provide an empirical calibration that relates the single temperature $T_{\rm e, [NII]}$ with metallicities derived from multiple ionization zones. 
\revtwo{Modeling \hii\ regions with multiple temperature zones is a classic technique (see, e.g., \citealt{Berg2020ApJ...893...96B}), however by using a single indicative \te, a significantly larger number of \hii\ regions can serve as a benchmark for cross-comparison with different strong-line calibrations.} From our sample of 556 \hii\ regions with \oii\ detections, 66 also have $T_{\rm e, [NII]}$ from \cite{Brazzini2024A&A...691A.173B} (adapting the S/N threshold as described in \citealt{Kreckel2025A&A...703A..42K}). Figure \ref{fig:strong_line_comp} shows a comparison of the metallicities resulting from \citetalias{Delgado2023a}~[\temp{N}{II}] in these to metallicities calculated for our strong line calibrations. 
We assess the level of agreement by both a Spearman $\rho$ and Pearson $r$ statistic (upper left), all with $p$-values less than 0.01. Most show a strong monotonic correlation ($\rho > 0.75$), although few show a strong linear correlation ($r >$ 0.5). \revone{The DESIRED calibrations (bottom row) show particularly strong correlations ($\rho > 0.75$), although we note that these calibrations are also heavily dependent on \te\nii\ for their termination of 12+log(O/H).} 
\citetalias{P05}~[R23] shows a significantly weaker correlation. Only \citetalias{Dopita2016}~[N2S2] approximates the 1-to-1 line, although not with a linear relation. \citetalias{Pilyugin2016}-[Scal] and \citetalias{Pilyugin2016}-[Rcal] along with \citetalias{P05}~[R23] all return systematically lower values, while \citetalias{Kewley2002}~[N2O2] and \citetalias{Kobulunicky2004}~[R23] return systematically higher values. 
Unfortunately, this \hii\ region catalog does not span a wide range in metallicities, with none in the low-metallicity regime (nothing below 8.2 in any diagnostic), and it remains difficult to draw strong conclusions on which calibration should be preferred.

Another possible indicator for the relative accuracy of a given calibration comes from an examination of the scatter in the radial metallicity gradient, $\sigma$(O/H). \cite{Kreckel2019} and \cite{Groves2023} noted that the \citetalias{Pilyugin2016}-[Scal] shows a remarkably small scatter, approaching the uncertainties associated with the line flux measurements. Here, we note that \citetalias{Pilyugin2016}-[Rcal] returns even lower scatter, 0.022 to 0.027 dex, in our three galaxies. However, many calibrations show low $<$0.04~dex scatter including \citetalias{Kewley2002}~[N2O2], \citetalias{Kobulunicky2004}~[R23] and \citetalias{Pilyugin2016}-[Scal].  \revone{This low $<$0.04~dex scatter is also seen for the \citetalias{RosalesOrtega2026}~[N2S2Ha] calibration, although the other DESIRED calibrations show scatter closer to $\sim$0.05~dex.}


\begin{figure*}
\centering
  \includegraphics[width=0.8\textwidth]{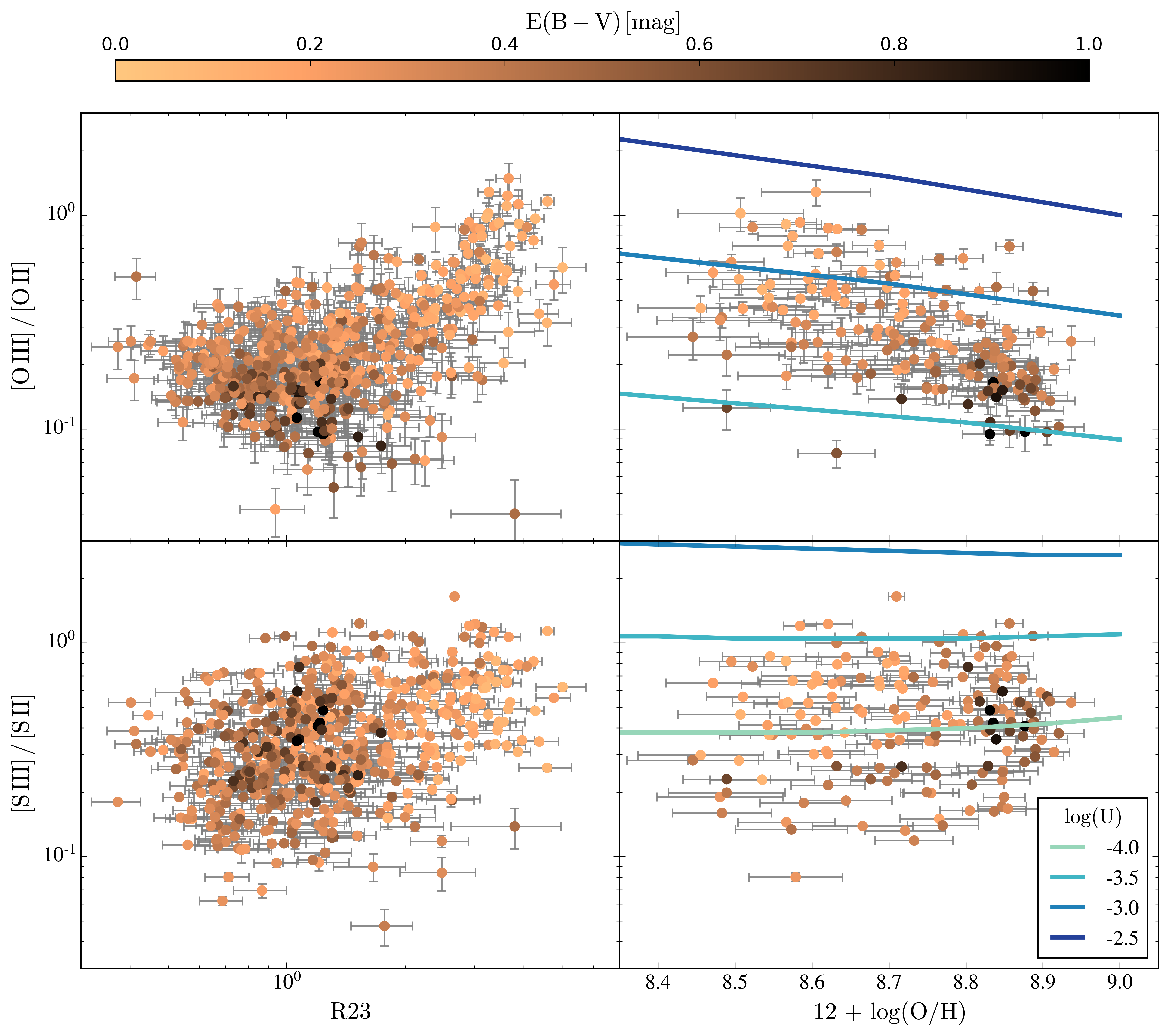}
  \caption{Each of the two line ratios, \oiii/\oii\ and \siii/\sii, as a function of R23 (left, a common metallicity diagnostic) \revone{and metallicity (right, using the \citetalias{Delgado2023a} \te\nii\ calibration).} All points are further colored by E(B-V), as measured from the Balmer decrement. Photoionization model lines \citep{Kewley2019} trace predicted values at constant ionization parameter (log $U$). }
  \label{fig:ion_param2}
\end{figure*}

\begin{figure}
    \centering
    \includegraphics[width=0.99\linewidth]{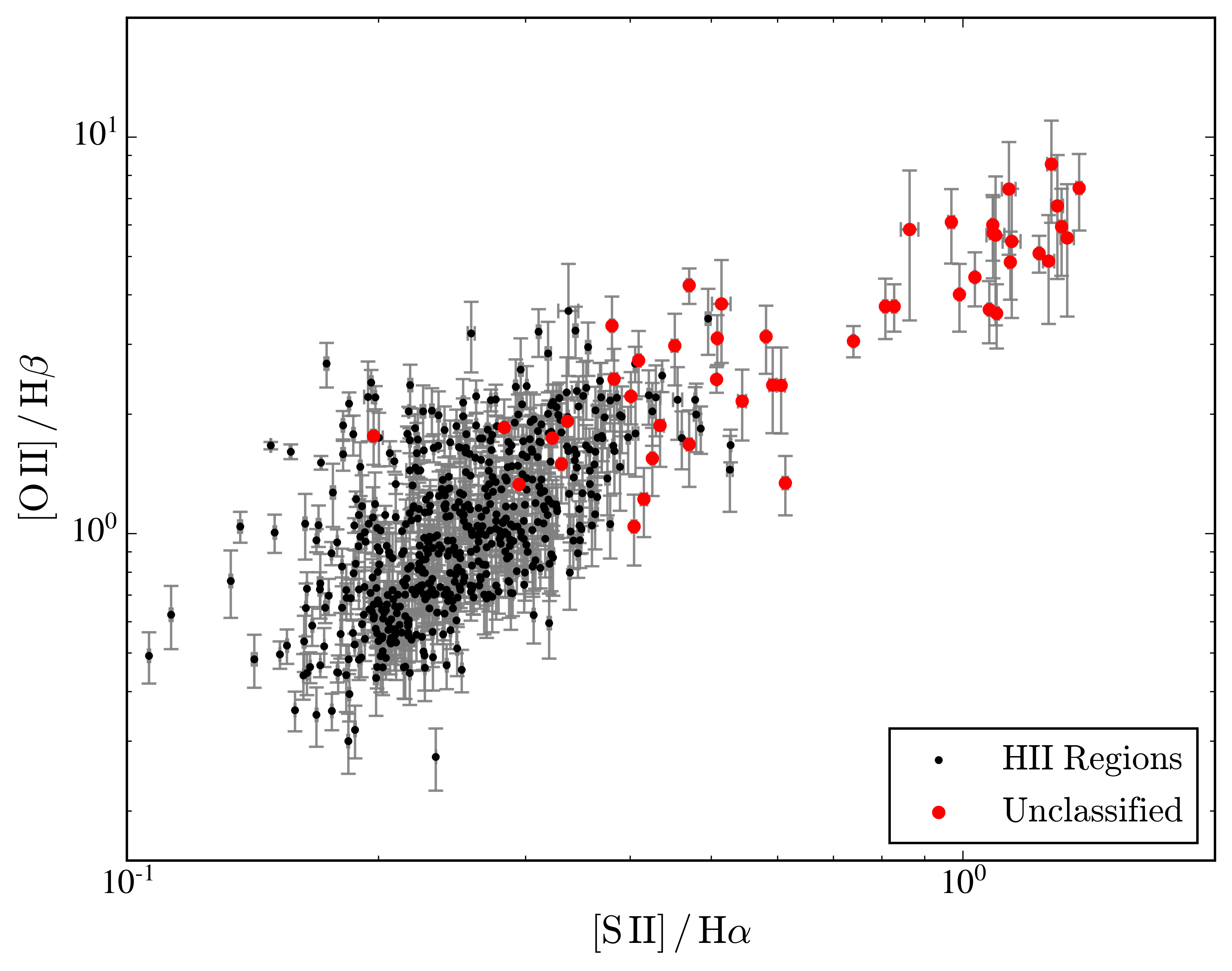}
    \caption{\oii/\hb\ as a function of \sii/\ha, comparing all \hii\ regions (black) with the 48 unclassified objects (red). Most exhibit high \sii/\ha, suggesting shock excitation, and about half overlap with known SNRs \citep{Li2024A&A...690A.161L}. }
    \label{fig:snrs}
\end{figure}

\subsection{Comparison of ionization parameter tracers}
\label{subsec:q_results}

\revtwo{Photoionization models predict that the \oiii/\oii\ ratio is sensitive to the ionization parameter, but also has a secondary dependence on metallicity and on the hardness of the ionizing radiation field \citep{Vilchez1988,Kewley2002}. In contrast, \siii/\sii\ is expected to provide a more direct tracer of the ionization parameter because it is less sensitive to metallicity than \oiii/\oii \citep[e.g.,][]{Dors2011}. This difference arises because \oiii/\oii\ is sensitive to changes in both the electron temperature of the gas and the hardness of the ionizing radiation field, which affect the relative strengths of [OIII] and [OII] \citep[e.g.,][]{Kewley2013, Delgado2023b}.} This assumption that \siii/\sii\ is a more direct tracer of ionization parameter was recently challenged by \cite{Garner2025arXiv251003144G}. 

In Figure \ref{fig:ion_param}, it is clear that while there is an overall positive correlation, these two line ratios are not tightly correlated with each other. In fact, individual galaxies (e.g., NGC~3627) even show an anti-correlation. 
To explore if the scatter could be driven by changes in metallicity, we plot each indicator against the R23 diagnostic as a proxy (Figure \ref{fig:ion_param2}), to keep our comparison grounded in observables. R23 is double valued, however, all our regions are high metallicity and expected to sit on the upper branch (see Figure \ref{fig:cornerplot}). We note that the results are similar when metallicities estimated from strong line calibrations are used. \revone{For an independent constraint on metallicity, we also show each indicator as a function of the \citetalias{Delgado2023a} \temp{N}{II} calibration, for which \oiii/\oii\ shows a modest trend while \siii/\sii\ does not. Photoionization models at fixed ionization parameter (log $U$) are also included \citep{Kewley2019}.} We further color all points by their E(B-V), as the wide wavelength separation in both line pairs may lead to reddening correction errors and contribute to the scatter. 
\revtwo{We find that \oiii/\oii\ correlates with both R23 (as well as metallicity) and E(B-V), such that more metal-poor regions, characterized by higher R23 and lower E(B-V), show systematically higher \oiii/\oii, consistent with the predictions of \cite{Kewley2002}. In contrast, \siii/\sii\ shows no clear trends with either quantity, suggesting that it provides a more direct tracer of variations in the ionization parameter, as predicted by photoionization models.}

\subsection{Outliers and Azimuthal variations}

In general, the scatter with respect to the radial metallicity gradient in all galaxies is small \revone{($<$0.1 dex),} reflecting that these inner parts of the galaxies can be well modeled by a simple linear fit. Outliers to low metallicity seen in some calibrations (e.g., \citetalias{P05}~[R23]) are generally not reflected in alternate diagnostics, which suggests that other changes in local condition (e.g., extinction, gas density, ionizing source,  hardness of the ionizing radiation) may be driving these outliers rather than true changes in metallicity. 


Interestingly, the radial metallicity gradients in \citetalias{Pilyugin2016}~[Scal] and \citetalias{Dopita2016}~[N2S2] show a secondary correlation with \siii/\sii, which suggests the lack of \oii\ in these calibrations may result in uncorrected ionization parameter dependencies. \revtwo{This was interpreted as a physical effect by \cite{Kreckel2019}}, however the lack of trend in other calibrations is troubling. This correlation between offset from the radial metallicity gradient ($\Delta$(O/H)) and ionization parameter was also seen\revtwo{ by \cite{Grasha2022ApJ...929..118G} when} using a metallicity based on the N2O2 diagnostic and ionization parameter based on \oiii/\oii, but we do not recover this in our three galaxies.

Of particular interest is the search for evidence of pristine hydrogen gas accretion, which may be signaled by a decrease in O/H \citep{Hwang2019ApJ...872..144H}, particularly if it is seen in relation to other relative abundances such as N/O \citep{Egorova2026A&A...705A..45E}. \revtwo{However, variations in dust depletion could also affect the observed gas-phase O/H, since a fraction of oxygen may be locked into dust grains \citep{Peimbert2010}. Therefore, localized O/H depressions should be interpreted cautiously and, where possible, compared against dust-sensitive quantities such as E(B-V), dust-to-gas ratio, or depletion-sensitive abundance ratios. Regardless, w}e see no evidence for any outlying \hii\ regions with unusually high N/O. Similarly, in the \citetalias{Kewley2002} metallicity calibration, which is determined using the N2O2 diagnostic and is therefore sensitive to variations in N/O, \revtwo{produces a tight radial metallicity relation (Figure~\ref{fig:strong_line_grad}). Together, these results show no significant evidence for ongoing or recent pristine gas accretion, although they do not rule out accretion at earlier times if the accreted gas has since mixed with the surrounding ISM.}

The current analysis of some of these trends is limited by our focus in this work on existing strong line calibrations. A more careful search and exploration of cross-diagnostic correlations requires careful treatment, such as simultaneously modeling multiple properties, such as N/O, O/H, and ISM pressure, as can be done with codes like HII-CHI-Mistry \citep{Perez-Montero2014MNRAS.441.2663P}, or was explored using Bayesian modeling techniques \citep{Bresolin2025MNRAS.539..755B}.

\subsection{Unclassified \oii\ Detections}

We note that 47 of the \oii\ detected nebulae (8\%) are unclassified (e.g. not classified as \hii\ regions). \revtwo{20 of these overlap with known supernova remnants (SNRs) from \cite{Li2024A&A...690A.161L} and 11 SNRs from \cite{Kravstov2025}.}  This is similar to the statistics on unclassified objects across the full Nebular Catalog, where $\sim$20\% of objects are `unclassified', and $\sim$5\% of the Nebular Catalog can be cross-matched to SNRs and SNR candidates. These all generally appear as outliers in an \oii/\hb\ vs. \sii/\ha\ diagnostic diagram (Figure \ref{fig:snrs}), exhibiting both high \sii/\ha\ (typical for SNRs) and high \oii/\hb. \revtwo{The separation of these sources in Figure \ref{fig:snrs} demonstrates that \oii\ provides useful leverage for distinguishing nebular environments in the ISM. In future work, SNR locations will be used to investigate the effectiveness of the \oii-based diagnostic and how local ISM environment may affect temperature inhomogeneities in nearby \hii\ regions (see, e.g., \citealt{Delgado2023a,RickardsVaught2024ApJ...966..130R}).}

\section{Conclusions}
\label{sec:conclusions}

We present an analysis of SITELLE \oii\ observations in five nearby galaxies, all part of the PHANGS-MUSE survey. We extract line fluxes for 604 regions, previously identified in the \cite{Groves2023} Nebular catalog. 556 of these are classified as \hii\ regions, and can be used to explore changes in the metallicity and ionization parameter. Three of the five galaxies have a large ($>$100) sample of \hii\ regions per galaxy, and radial trends can also be explored.

We release a catalog of all measured line fluxes, resulting in a combined database (together with \revtwo{measurements from MUSE}) that includes both \oii\ and \siii, something not always available 
in the literature. 
In combination with the auroral line covered by MUSE, particularly \oii$\lambda\lambda$7320,7330, this is also well suited for multi-zone temperature modeling of these \hii\ regions \citep{Habjan2026b}. Our key results are listed below.

Among the different strong-line calibrations examined in this work, we find long-standing offsets and calibration-dependent variations. Notably, the \citetalias{Pilyugin2016}-[Rcal] calibration yields very small scatter, $\sigma({\rm O/H}) = 0.022$--$0.027$~dex, about the radial metallicity gradients. While this may suggest efficient mixing in the ISM, the extent to which this scatter reflects intrinsic abundance variations rather than a methodological floor remains uncertain.

\begin{itemize}
    \item  Among the different strong line calibrations derived in this work, long-standing offsets and variations are demonstrated. Remarkably, the \citetalias{Pilyugin2016}-[Rcal] shows very small scatter (0.022-0.027~dex) in the radial metallicity gradients\revtwo{. While this may suggest efficient mixing in the ISM, the extent to which this scatter reflects intrinsic abundance variations rather than a methodological floor remains uncertain.} 
    \item  \siii/\sii\ shows no strong secondary correlation with metallicity (via R23) or extinction, in contrast to \oiii/\oii, which suggests \siii/\sii\ is a more direct and robust tracer of ionization parameter. 
    \item We find no outliers in O/H or in N/O, although our focus on the inner disk region implies efficient mixing may quickly remove any such variations due to gas accretion or gas flows. 
    \item We identify \oii\ detections in 48 unclassified regions, many of which appear likely to be SNRs.

\end{itemize}

While our analysis focused on exploitation of existing strong line calibrations, a more detailed approach using  Bayesian inference \citep{Blanc2015ApJ...798...99B} or machine learning \citep{Ho2019MNRAS.485.3569H} could be used to improve the simultaneous fitting of multiple physical conditions. 
It is clear that the addition of \oii\ line fluxes improves our ability to infer metallicities from strong lines alone, and with the upcoming BlueMUSE in synergy with MUSE, we expect forthcoming rich multi-line datasets to provide much-needed breakthroughs in the robust interpretation of emission line diagnostics.

\section*{Acknowledgements}

KK gratefully acknowledges funding from the Deutsche Forschungsgemeinschaft (DFG, German Research Foundation) in the form of an Emmy Noether Research Group (grant number KR4598/2-1, PI Kreckel) and the European Research Council’s starting grant ERC StG-101077573 (“ISM-METALS"). 

RSK acknowledges financial support from the ERC via Synergy Grant ``ECOGAL'' (project ID 855130) and from the German Excellence Strategy via the Heidelberg Cluster ``STRUCTURES'' (EXC 2181 - 390900948). In addition RSK is grateful for funding from the German Ministry for Economy and Energy (BMWE) in project ``MAINN'' (funding ID 50OO2206), and from DFG and ANR for project ``STARCLUSTERS'' (funding ID KL 1358/22-1). 

TGW gratefully acknowledges support from the UK ALMA Regional Centre (ARC) Node, which is supported by the Science and Technology Facilities Council grant number ST/Y004108/1.

HAP acknowledges support from the National Science and Technology Council of Taiwan under grant 113-2112-M-032-014-MY3.

OE acknowledges funding from the Deutsche Forschungsgemeinschaft (DFG, German Research Foundation) -- project-ID 541068876.

Table \ref{tbl:sample_observations} includes distances that were compiled or calculated by \citet{Anand2021}. The distance to NGC~628, NGC~3351, and NGC~3627 was calculated using the tip of the red giant branch (TRGB) method in \cite{Jacobs2009}, NGC~2835 with TRGB in \cite{Anand2021}, and NGC~4535 with Cepheid variable stars in \cite{Freedman2001}.

Based on observations obtained at the Canada-France-Hawai'i Telescope (CFHT) which is operated by the National Research Council of Canada, the Institut National des Sciences de l'Univers of the Centre National de la Recherche Scientifique of France, and the University of Hawai'i. CFHT is located on Maunakea on Hawai'i Island, a mountain of considerable cultural, natural, and ecological significance. Maunakea is a sacred site to Native Hawaiians, also known as Kānaka 'Ōiwi. We would like to thank the Canada-France-Hawai'i Telescope (CFHT) Operations and Software Groups for their contributions and diligence in maintaining observatory operations; the CFHT Astronomy Group for their observation coordination and data acquisition efforts; and the CFHT Finance \& Administration Group for their contributions to the management and administration of the observatory. Based on observations obtained with SITELLE, a joint project between Universit\'e Laval, ABB-Bomem, Universit\'e de Montreal, and the CFHT with funding support from the Canada Foundation for Innovation (CFI), the National Sciences and Engineering Research Council of Canada (NSERC), Fond de Recheche du Quebec - Nature et Technologies (FRQNT) and CFHT.

Based on observations collected at the European Southern Observatory under ESO programmes 094.C-0623 (PI: Kreckel), 095.C-0473,  098.C-0484 (PI: Blanc), 1100.B-0651 (PHANGS-MUSE; PI: Schinnerer), as well as 094.B-0321 (MAGNUM; PI: Marconi), 099.B-0242, 0100.B-0116, 098.B-0551 (MAD; PI: Carollo) and 097.B-0640 (TIMER; PI: Gadotti). 

This research has made use of several Python packages, namely the main \textsc{astropy} package \citep{Astropy+2013, Astropy+2018, Astropy+2022}, \textsc{numpy} \citep{Harris+2020} and \textsc{matplotlib} \citep{Hunter+2007}.

\bibliographystyle{mnras}
\bibliography{refs}

\newpage

\begin{appendix}

\section{Comparison with literature data}
\label{app:lit_validation}

\begin{figure*}[!h]
\centering
  \includegraphics[width=0.85\textwidth]{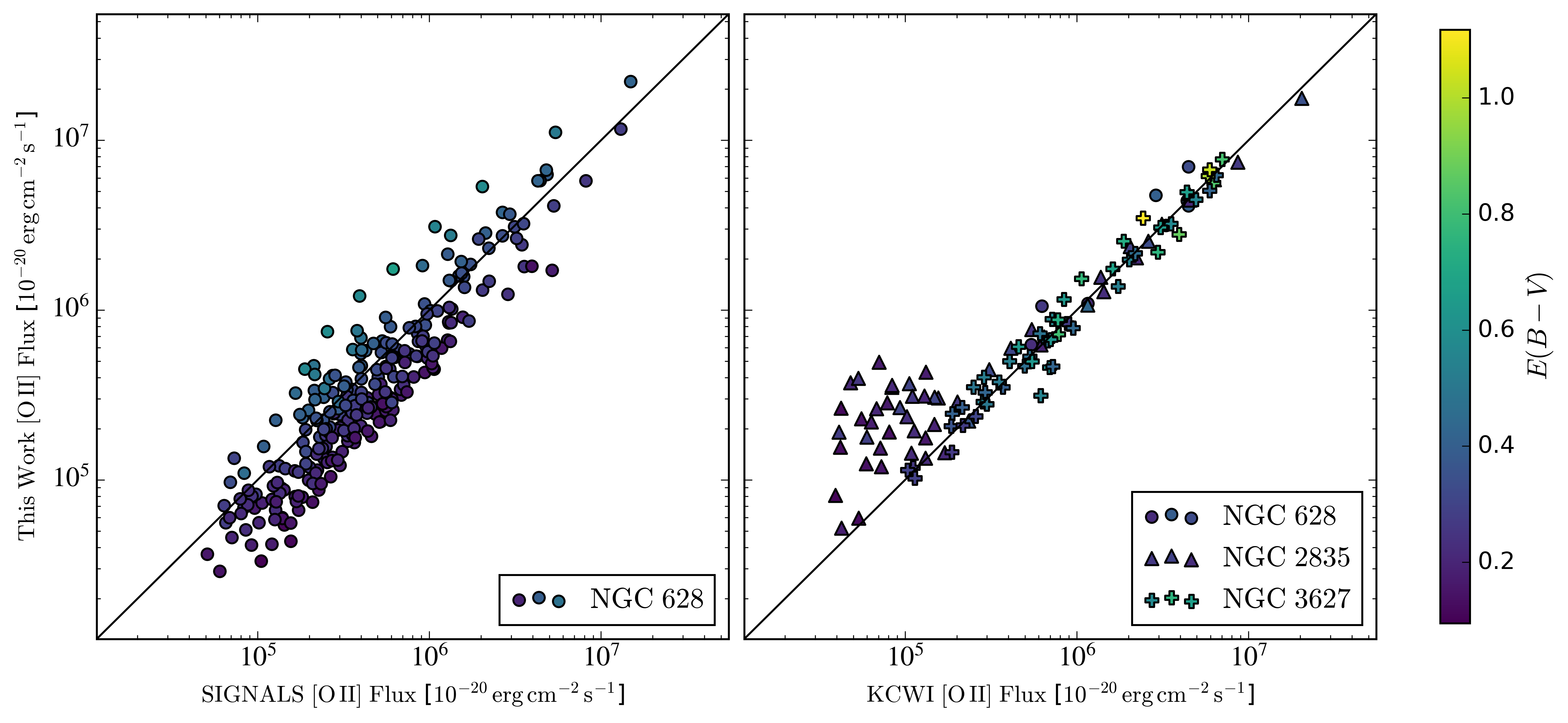}
  \caption{Comparison of extinction-corrected \oii $\lambda$3727 fluxes measured in this work with literature values from \cite{Rousseau-Nepton2018} for NGC~628 (left) and \cite{RickardsVaught2024ApJ...966..130R} for NGC~628, NGC~2835, and NGC~3627 (right). Points are colored by $E(B-V)$, and the solid line indicates one-to-one agreement.}
  \label{fig:lit_comp}
\end{figure*}

\oii\ measurements are available for \hii\ regions in three of our galaxies in the literature, and we compare here with those results to validate our line flux measurements. 
The first comparison was performed using \oii\ fluxes from the KCWI study of \cite{RickardsVaught2024ApJ...966..130R}, which included eight \hii\ regions in NGC~628, 53 \hii\ regions in NGC~2835\revtwo{, and 53 \hii\ regions in NGC~3627}. The SITELLE observations were reprojected to match the spatial and pixel dimensions of the KCWI observations. The \hii\ region mask developed in \cite{RickardsVaught2024ApJ...966..130R} was then used to extract \hii\ region spectra. 
The full comparison is shown in Figure \ref{fig:lit_comp}. 
At high fluxes, the \oii\ flux varies by 23.1\% in six NGC~628 regions, 8.9\% in 12 NGC~2835 regions\revtwo{, and 15.3\% in 19 NGC~3627 regions}. 
The agreement worsens below $10^{-14} \text{ erg}\text{ cm}^{-2}\text{ s}^{-1}$, with all regions below this threshold coming from NGC~2835 and NGC~3627. 

The second comparison was done with the SITELLE study by \cite{Rousseau-Nepton2018} for 271 regions in NGC~628. The fully analyzed \oii\ maps received from the authors \revtwo{required a small astrometric correction to align with} the Nebular Catalog \cite{Groves2023}\revtwo{; this correction was applied following the methodology described in Section~\ref{subsec:alignmet}}. \oii\ fluxes were then extracted using the segmentation from the Nebular catalog. Across these 271 regions, a percent error of 35.9\% was measured. The scatter in this comparison appears to be driven primarily by two main factors: (1) a dependence on $E(B-V)$ and (2) differences in the treatment of the diffuse ionized gas (DIG), as \cite{Rousseau-Nepton2018} included a DIG subtraction. Applying a similar DIG subtraction to our data resulted in a $\sim10\%$ difference in \oii\ fluxes, indicating that DIG subtraction accounts for part, but not all, of the observed difference. Future work should consider further studying the differences due to reddening and DIG contribution in more detail; this is particularly important for metallicity calibrations using \nii\ and {\sii}. \revtwo{Because the analysis presented here relies on line ratios that all use the $E(B-V)$ from the nebular catalog, rather than on combining \oii\ fluxes from \cite{Rousseau-Nepton2018} with optical line fluxes from the nebular catalog, this offset is unlikely to be significant bias in the strong line metallicity analysis. Nevertheless, the offset highlights the need for careful treatment of reddening corrections and DIG subtraction in future direct comparisons between SITELLE and MUSE measurements.}

\clearpage

\section{Summary information for strong line calibrations}
\label{app:strong_line_summary}

In this work, we consider a variety of strong line calibrations, as detailed in Table \ref{tbl:prescriptions}. A  brief summary of each is given below:

\begin{itemize}

\item \cite{Kewley2002}. Hereafter, KD02~[N2O2], provides multiple theoretical calibrations based on photoionization models. These models adopt a solar oxygen abundance of $\rm 12 + \log(O/H) = 8.93$. We focus on their calibration using the \nii/\oii\ line ratio. The expression \citetalias{Kewley2002}~[N2O2] is:

\begin{gather}
\label{eq:KD02}
\begin{aligned}
\rm 12 + \log\left(\rm O/H\right) = \log \left( 1.54020 + 1.26602 \times N2O2 + 0.167977 \times N2O2^2 \right) + 8.93,
\end{aligned}
\end{gather}

\noindent which is only valid for high metallicity systems (12+log(O/H)$>$8.6). This calibration is expected to correlate directly with N/O as it increases linearly with O/H due to secondary nitrogen production in intermediate-mass stars and be insensitive to the ionization parameter. It is proposed as one of the most reliable metallicity calibrations \citep{Kewley2019}. 

\item \cite{Kobulunicky2004}. Hereafter KK04~[R23], is a theoretical calibration that uses stellar evolution and photoionization model grids from \cite{Kewley2002}. \citetalias{Kobulunicky2004} adopt 
$12+\log({\rm O/H})_{\odot} = 8.72$, although the underlying 
\cite{Kewley2002} photoionization grids were originally constructed using $12+\log({\rm O/H})_{\odot} = 8.93$. In our implementation, we use Equations~A4--A6 from \cite{Kewley2008}, which are solved iteratively to determine the metallicity. For the first iteration, we estimate the ionization parameter using Equation~A4 with an initial metallicity of $12 + \log(\mathrm{O/H}) = 8.2$ for the lower branch, defined by N2O2 $< -1.2$, or $12 + \log(\mathrm{O/H}) = 8.7$ for the upper branch, defined by N2O2 $\geq -1.2$:

\begin{gather}
\label{eq:KK04_logq}
\begin{aligned}
\log q =
\frac{
32.81 - 1.153y^2
+ \left[12 + \log\left(\mathrm{O/H}\right)\right]
\left(-3.396 - 0.025y + 0.1444y^2\right)
}{
4.603 - 0.3119y - 0.163y^2
+ \left[12 + \log\left(\mathrm{O/H}\right)\right]
\left(-0.48 + 0.0271y + 0.02037y^2\right)
},
\end{aligned}
\end{gather}

\noindent where $y = \rm O32 = $~$\log$(\oiii/\oii). Then to calculate the metallicity, we use either the lower or upper branch:

\begin{gather}
\label{eq:KK04_lower}
\begin{aligned}
12 + \log\left(\mathrm{O/H}\right)_{\rm lower}
= 9.40 + 4.65x - 3.17x^2
- \log q \left(0.272 + 0.547x - 0.513x^2\right),
\end{aligned}
\end{gather}

\begin{gather}
\label{eq:KK04_upper}
\begin{aligned}
12 + \log\left(\mathrm{O/H}\right)_{\rm upper}
= 9.72 - 0.777x - 0.951x^2 - 0.072x^3 - 0.811x^4 \\
{} - \log q \left(
0.0737 - 0.0713x - 0.141x^2 + 0.0373x^3 - 0.058x^4
\right),
\end{aligned}
\end{gather}

\noindent where $x = \rm R23$. In each subsequent iteration, we recalculate the ionization parameter from Equation~\ref{eq:KK04_logq} using the metallicity from the previous iteration, then update the metallicity using Equation~\ref{eq:KK04_lower} for $12 + \log(\mathrm{O/H}) \leq 8.4$ or Equation~\ref{eq:KK04_upper} for $12 + \log(\mathrm{O/H}) > 8.4$. We consider the metallicity converged after five iterations.

\item \cite{P05}. Hereafter PT05~[R23], is an empirical calibration that updates the calibration from \cite{Pilyugin2001} using a larger sample of \hii\ regions that makes use of R23. These direct metallicities are determined using \te derived from the \oiii$\lambda$4363 auroral line. We use the upper and lower branches determined in \cite{Kewley2008} using the \nii/\oii\ line ratio. The lower branch is defined as:

\begin{gather}
\label{eq:P05_lower}
\begin{aligned}
12 + \log\left(\mathrm{O/H}\right)_{\rm lower}
= \rm
\frac{
R23 + 106.4 + 106.8 \times P - 3.40 \times P^2
}{
17.72 + 6.60 \times P + 6.95 \times P^2 - 0.302 \times R23
},
\end{aligned}
\end{gather}

\noindent And the upper branch as:

\begin{gather}
\label{eq:P05_upper}
\begin{aligned}
12 + \log\left(\mathrm{O/H}\right)_{\rm upper}
= \rm 
\frac{
R23 + 726.1 + 842.2 \times P + 337.5 \times P^2
}{
85.96 + 82.76 \times P + 43.98 \times P^2 + 1.793 \times R23
}.
\end{aligned}
\end{gather}

\noindent where \(\mathrm{P} = ([\mathrm{O\,III}]\,\lambda\lambda4959,5007/\mathrm{H}\beta)/\mathrm{R23}\).

\item \citet{Pilyugin2016}. Hereafter PG16, introduces three different empirical strong line methods that are calibrated using direct $T_{e}$ metallicities determined in \cite{Pilyugin2016}. Since \citetalias{Pilyugin2012} calibrates these directly to $T_e$ abundances there is no adopted solar value. The R calibration (hereafter, \citetalias{Pilyugin2016}-[Rcal]) uses $R_2 =$ \oii $\lambda$3727~/~H$\beta$, $R_3 =$ \oiii $\lambda\lambda$4959, 5007/H$\beta$, and $N_2 =$~\nii$\lambda\lambda$6548, 6584/H$\beta$ line ratios. 
This calibration is separated into lower and upper branches using N2, where the lower branch is defined by $\log(\mathrm{N_2}) < -0.6$ and the upper branch is defined by $\log(\mathrm{N_2}) \geq -0.6$. The lower branch is given by:

\begin{gather}
\label{eq}
\begin{aligned}
12 + \log\left(\mathrm{O/H}\right)_{\rm lower}
= 7.932 + 0.944 \log\left(\frac{R_3}{R_2}\right)
+ 0.695 \log\left(\mathrm{N_2}\right) \\
{} + \left[
0.970 - 0.291 \log\left(\frac{R_3}{R_2}\right) + 0.019 \log\left(\mathrm{N_2}\right)
\right]\log\left(R_2\right).
\end{aligned}
\end{gather}

\noindent The upper branch is given by:

\begin{gather}
\label{eq:PG16_Rcal_upper}
\begin{aligned}
12 + \log\left(\mathrm{O/H}\right)_{\rm upper}
= 8.589 + 0.022 \log\left(\frac{R_3}{R_2}\right)
+ 0.399 \log\left(\mathrm{N_2}\right) \\
{} + \left[
-0.137 + 0.164 \log\left(\frac{R_3}{R_2}\right)
+ 0.589 \log\left(\mathrm{N_2}\right)
\right]\log\left(R_2\right).
\end{aligned}
\end{gather}

\item \citetalias{Pilyugin2016} derives a sulfur-dependent metallicity called the S calibration, hereafter \citetalias{Pilyugin2016}-[Scal]. The same metallicities from the \citetalias{Pilyugin2016}~[Rcal] are used to calibrate this empirical relation using the $S_2 =$ \sii$\lambda\lambda$6717, 6731 /H$\beta$, $R_3$ and $N_2$ line ratios. A lower branch is defined by $\log(\mathrm{N_2}) < -0.6$ and an upper branch is defined by $\log(\mathrm{N_2}) \geq -0.6$. The lower branch is given by:

\begin{gather}
\label{eq:PG16_Scal_lower}
\begin{aligned}
12 + \log\left(\mathrm{O/H}\right)_{\rm lower}
= 8.072 + 0.789 \log\left(\frac{R_3}{S_2}\right)
+ 0.726 \log\left(\mathrm{N_2}\right) \\
{} + \left[
1.069 - 0.170 \log\left(\frac{R_3}{S_2}\right)
+ 0.022 \log\left(\mathrm{N_2}\right)
\right]\log\left(S_2\right).
\end{aligned}
\end{gather}

\noindent The upper branch is given by:

\begin{gather}
\label{eq:PG16_Scal_upper}
\begin{aligned}
12 + \log\left(\mathrm{O/H}\right)_{\rm upper}
= 8.424 + 0.030 \log\left(\frac{R_3}{S_2}\right)
+ 0.751 \log\left(\mathrm{N_2}\right) \\
{} + \left[
-0.349 + 0.182 \log\left(\frac{R_3}{S_2}\right)
+ 0.508 \log\left(\mathrm{N_2}\right)
\right]\log\left(S_2\right).
\end{aligned}
\end{gather}

\noindent This empirical calibration has been used commonly in PHANGS-MUSE studies \citep[e.g.,][]{Kreckel2019}.

\item \cite{Dopita2016}. Hereafter D16~[N2S2], collects derived N/O - O/H relationships from the literature, which are vital to the photoionization models that \citetalias{Dopita2016} uses to create this calibration. These models adopt local Galactic concordance abundances with $12+\log({\rm O/H})=8.77$. This combined strong line method makes use of the \nii$\lambda$6584, H$\alpha$, and \sii$\lambda\lambda$6716, 6731 emission lines. In particular: 

\begin{gather}
\label{eq:D16_y}
\begin{aligned}
y =
\log \left(
\frac{[\mathrm{N\,II}]\,\lambda6583}
{[\mathrm{S\,II}]\,\lambda6717 + [\mathrm{S\,II}]\,\lambda6731}
\right)
+ 0.264 \log \left(
\frac{[\mathrm{N\,II}]\,\lambda6583}
{\mathrm{H}\alpha}
\right),
\end{aligned}
\end{gather}

\noindent which is used to define:

\begin{gather}
\label{eq:D16_met}
\begin{aligned}
12 + \log\left(\mathrm{O/H}\right)
= 8.77 + y + 0.45\left(y + 0.3\right)^5.
\end{aligned}
\end{gather}

\noindent The \citetalias{Dopita2016}~[N2S2] metallicity calibration is valid until 12 + log(O/H) $\sim$ 9.05.

\item \cite{RosalesOrtega2026}. Hereafter RO26, present the DEep Spectra of Ionised REgions Database (DESIRED) empirical strong line calibrations, derived from a large compilation of deep \hii\ region and star forming galaxy spectra with direct electron temperature abundances. Since \citetalias{RosalesOrtega2026} calibrates these directly to $T_e$ abundances there is no adopted solar value; each calibration is written as a polynomial:

\begin{gather}
\label{eq:RO26}
\begin{aligned}
12 + \log\left(\mathrm{O/H}\right)
= a_0 + a_1x + a_2x^2 + a_3x^3.
\end{aligned}
\end{gather}

In Table~\ref{tbl:RO26_coefficients} we show the logarithmic line ratio $x$, the validity range of each calibration in line-ratio space, and the standard deviation of residuals from the \te derived abundances ($\sigma_{\rm cal}$) in \citetalias{RosalesOrtega2026}. The \citetalias{RosalesOrtega2026}~[R23] and \citetalias{RosalesOrtega2026}~[$\hat{R}$] calibrations make use of lower and upper branch coefficients, as indicated in Table~\ref{tbl:RO26_coefficients}.

\begin{table*}[!h]
\centering
\footnotesize
\caption{\citetalias{RosalesOrtega2026} strong line calibration coefficients.}
\label{tbl:RO26_coefficients}
\begin{tabular}{c|c|c|cccc|c}
\toprule
Index $x$ & Branch & Validity & $a_0$ & $a_1$ & $a_2$ & $a_3$ & $\sigma_{\rm cal}$ \\
\hline \hline
R23 & upper & $[-0.20,\,1.03]$ & 8.67 & $-0.35$ & 0.00 & $-0.28$ & 0.17 \\
R23 & lower & $[0.35,\,1.03]$ & 6.82 & 0.49 & 0.05 & 0.54 & 0.19 \\
$\hat{R}$ & upper & $[-1.14,\,0.79]$ & 8.45 & $-0.32$ & $-0.18$ & $-0.10$ & 0.16 \\
$\hat{R}$ & lower & $[-0.20,\,0.79]$ & 7.10 & 0.70 & 0.37 & 0.21 & 0.16 \\
N2O2 & -- & $[-1.87,\,0.35]$ & 8.61 & 0.38 & 0.18 & 0.20 & 0.25 \\
N2S2 & -- & $[-0.89,\,0.50]$ & 8.37 & 0.53 & 0.05 & 0.92 & 0.24 \\
N2S2Ha & -- & $[-1.48,\,0.40]$ & 8.47 & 0.55 & 0.47 & 0.46 & 0.22 \\
\bottomrule
\end{tabular}
\end{table*}

\end{itemize}

\begin{figure*}[!h]
    \centering
    \includegraphics[width=0.9\textwidth]{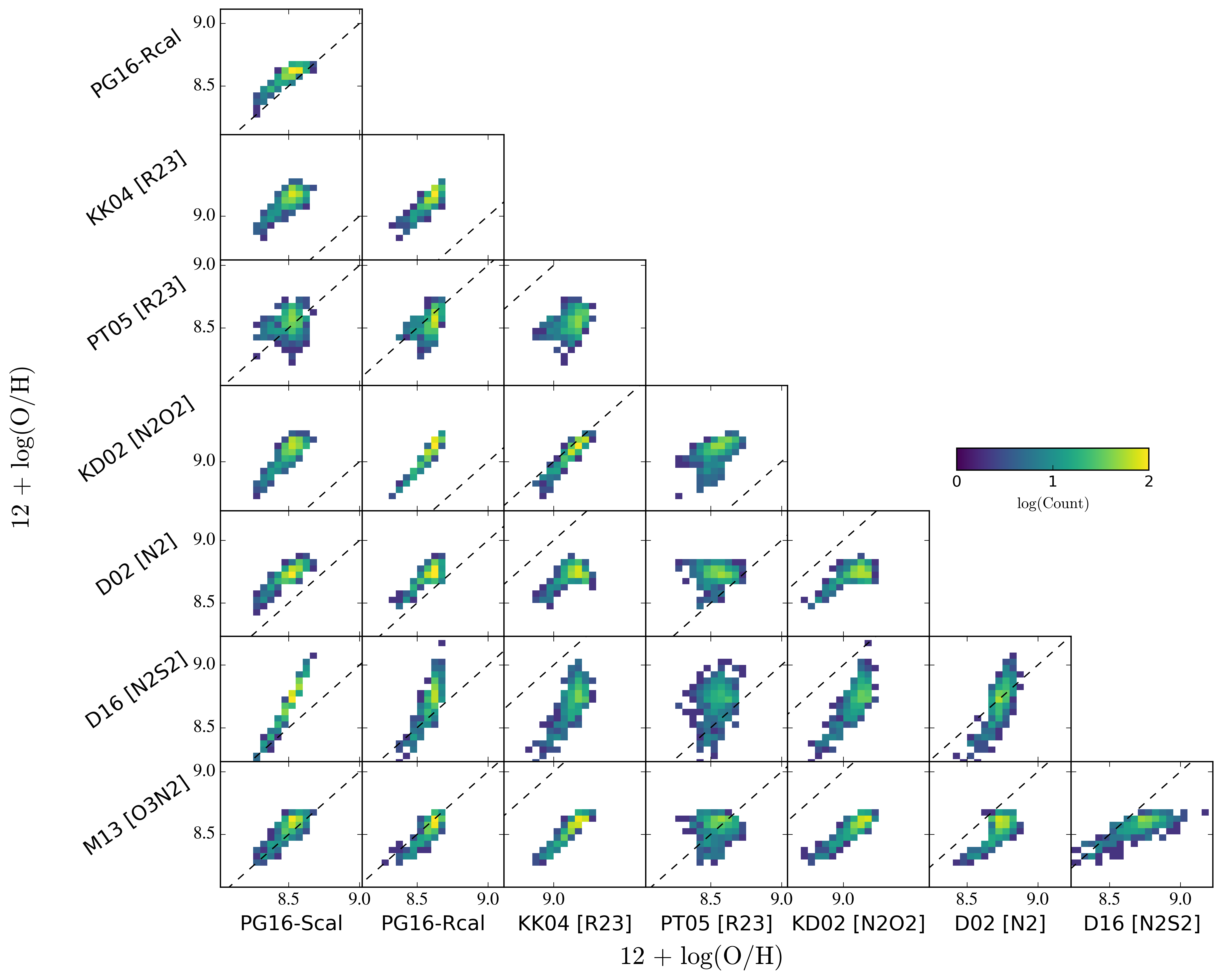}
      \caption{2D histograms showing a comparison of eight different metallicity calibrations (see text), as applied to 556 \hii\ regions. The intensity indicates the number of \hii\ regions in each bin. The 1-to-1 line is shown in each panel (dashed), to make both the correlations and the absolute offsets between calibrations easier to identify. }
    \label{fig:cornerplot}
\end{figure*}

\noindent For completeness, we also considered two additional calibrations that focused on diagnostic line ratios shown in \cite{Groves2023} to exhibit significant scatter and deviations from other calibrations:

\begin{itemize}

\item \cite{D02}. Hereafter D02~[N2], is a semi-empirical strong line calibration. At low metallicity, the reference oxygen abundances
are based on \te derived abundances using \oiii$\lambda$4363, while at high metallicity strong line abundances from \cite{M91} and \cite{Diaz2000} are used. D02 take the solar oxygen abundance to be $12+\log({\rm O/H})_{\odot}=8.91$.. The D02 calibration uses N2 and is defined as:

\begin{gather}
\label{eq:D02_met}
\begin{aligned}
\rm 12 + \log\left(\mathrm{O/H}\right)
= 9.12 + 0.73\times \,N2.
\end{aligned}
\end{gather}

\item \cite{Marino2013}. Hereafter M13~[O3N2], calibrates two different empirical strong line methods using a sample of CEL derived metallicities with directly measured \temp{O}{III}, \temp{N}{II}, and \temp{S}{III}. Since this is an empirical calibration from direct \te derived abundances, there is no normalizing solar oxygen abundance. We consider only the second calibration from M13, M13-O3N2, which provides an empirical calibration using the line ratio: 

\begin{gather}
\label{eq:M13_O3N2}
\begin{aligned}
\rm O3N2 =
\log \left[
\frac{[\mathrm{O\,III}]\,\lambda5007}
{\mathrm{H}\beta}
\times 
\frac{\mathrm{H}\alpha}
{[\mathrm{N\,II}]\,\lambda6583}
\right],
\end{aligned}
\end{gather}

\noindent and the calibration is defined as:

\begin{gather}
\label{eq:M13_O3N2_met}
\begin{aligned}
\rm 12 + \log\left(\mathrm{O/H}\right)
= 8.533 - 0.214\,\times O3N2.
\end{aligned}
\end{gather}

\end{itemize}

\noindent A complete comparison of all metallicities resulting from these calibrations is shown in Figure \ref{fig:cornerplot}.

\end{appendix}

\end{document}

%% file: authors.tex
\author{Eric~Habjan$^{1,2}$\orcidlink{0009-0003-9547-0952}} %
\author{Kathryn~Kreckel$^{3}$\orcidlink{0000-0001-6551-3091}}
\author{Christopher~Faesi$^{2}$\orcidlink{0000-0001-5310-467X}}
\author{Francesco~Belfiore$^{4,5}$\orcidlink{0000-0002-2545-5752}}
\author{Brent Groves$^{6}$\orcidlink{0000-0002-9768-0246}}
\author{J.~Eduardo~M{\'e}ndez-Delgado$^{7}$\orcidlink{0000-0002-6972-6411}}
\author{Ryan~J.~Vaught$^{8}$\orcidlink{0000-0001-9719-4080}}
\author{Laurie~Rousseau-Nepton$^{9, 10, 11}$}
\author{Fabian~Scheuermann$^{3}$\orcidlink{0000-0003-2707-4678}}
\author{F.~Fabi{\'a}n~Rosales-Ortega$^{12}$\orcidlink{0000-0002-3642-9146}}
\author{Hsi-An~Pan$^{13}$\orcidlink{0000-0002-1370-6964}}
\author{Daniel~A.~Dale$^{14}$\orcidlink{0000-0002-5782-9093}}
\author{Thomas~G.~Williams$^{15}$\orcidlink{0000-0002-0012-2142}}
\author{Ralf~S.\ Klessen$^{16,17}$\orcidlink{0000-0002-0560-3172}}
\author{Oleg~V.~Egorov$^{3}$\orcidlink{0000-0002-4755-118X}}  
\author{Timo~Kravtsov$^{18,19}$\orcidlink{0000-0003-0955-9102}}
\author{Amirnezam~Amiri$^{20,21}$\orcidlink{0000-0002-8553-1964}}
\author{Kathryn~Grasha$^{22}$\orcidlink{0000-0002-3247-5321}}
\author{Eric~Emsellem$^{4,23}$\orcidlink{0000-0002-6155-7166}}

\email{habjan.e@northeastern.edu}

\affiliation{$^{1}$Department of Physics, Northeastern University, 110 Forsyth St, Boston, MA 02115}
\affiliation{$^{2}$University of Connecticut, Department of Physics, 196A Auditorium Road, Unit 3046, Storrs, CT 06269, USA}
\affiliation{$^{3}$Astronomisches Rechen-Institut, Zentrum f\"{u}r Astronomie der Universit\"{a}t Heidelberg, M\"{o}nchhofstr. 12-14, D-69120 Heidelberg, Germany}
\affiliation{$^{4}$European Southern Observatory, Karl-Schwarzschild Stra{\ss}e 2, D-85748 Garching bei M\"{u}nchen, Germany}
\affiliation{$^{5}$INAF -- Osservatorio Astrofisico di Arcetri, Largo E. Fermi 5, I-50157 Firenze, Italy}
\affiliation{$^{6}$International Centre for Radio Astronomy Research, University of Western Australia, 7 Fairway, Crawley, 6009 WA, Australia}
\affiliation{$^{7}$Universidad Nacional Aut\'onoma de M\'exico, Instituto de Astronom\'ia, AP 70-264, CDMX 04510, M\'exico}
\affiliation{$^{8}$Space Telescope Science Institute, 3700 San Martin Drive, Baltimore, MD 21218, USA}
\affiliation{$^{9}$Dunlap Institute of Astronomy and Astrophysics, University of Toronto, 50 St. George St, Toronto, ON, M5S 3H4, Canada}
\affiliation{$^{10}$Department of Astronomy \& Astrophysics at the University of Toronto, 50 St. George St, Toronto, ON, M5S 3H4, Canada}
\affiliation{$^{11}$Canada-France-Hawaii Telescope, 65-1238 Mamalahoa Hwy, Kamuela, Hawaii 96743, USA}
\affiliation{$^{12}$Instituto Nacional de Astrof{\'i}sica, {\'O}ptica y Electr{\'o}nica (INAOE SECIHTI), Luis E. Erro 1, 72840, Tonantzintla, Puebla, M{\'e}xico}
\affiliation{$^{13}$Department of Physics, Tamkang University, No.151, Yingzhuan Road, Tamsui District, New Taipei City 251301, Taiwan}
\affiliation{$^{14}$Department of Physics and Astronomy, University of Wyoming, Laramie, WY 82071, USA}
\affiliation{$^{15}$UK ALMA Regional Centre Node, Jodrell Bank Centre for Astrophysics, Department of Physics and Astronomy, The University of Manchester, Oxford Road, Manchester M13 9PL, UK}
\affiliation{$^{16}$Universit\"{a}t Heidelberg, Zentrum f\"{u}r Astronomie, Institut f\"{u}r Theoretische Astrophysik, Albert-Ueberle-Str.\ 2, 69120 Heidelberg, Germany}
\affiliation{$^{17}$Universit\"{a}t Heidelberg, Interdisziplin\"{a}res Zentrum f\"{u}r Wissenschaftliches Rechnen, Im Neuenheimer Feld 225, 69120 Heidelberg, Germany}
\affiliation{$^{18}$Tuorla Observatory, Department of Physics and Astronomy, University of Turku, 20014Turku, Finland}
\affiliation{$^{19}$Finnish Centre for Astronomy with ESO (FINCA), University of Turku, Vesilinnantie 5, Quantum 20014 Turku, Finland}
\affiliation{$^{20}$School of Astronomy, Institute for Research in Fundamental Sciences (IPM), Tehran, P.O. Box 19395-5531, Iran}
\affiliation{$^{21}$Department of Physics, University of Arkansas, 226 Physics Building, 825 West Dickson Street, Fayetteville, AR 72701, USA}
\affiliation{$^{22}$Research School of Astronomy and Astrophysics, Australian National University, Canberra, ACT 2611, Australia}
\affiliation{$^{23}$Univ Lyon, Univ Lyon1, ENS de Lyon, CNRS, Centre de Recherche Astrophysique de Lyon UMR5574, F-69230 Saint-Genis-Laval France}

%% file: refs.bib
@ARTICLE{2020A&A...641A..28W,
       author = {{Weilbacher}, Peter M. and {Palsa}, Ralf and {Streicher}, Ole and {Bacon}, Roland and {Urrutia}, Tanya and {Wisotzki}, Lutz and {Conseil}, Simon and {Husemann}, Bernd and {Jarno}, Aur{\'e}lien and {Kelz}, Andreas and {P{\'e}contal-Rousset}, Arlette and {Richard}, Johan and {Roth}, Martin M. and {Selman}, Fernando and {Vernet}, Jo{\"e}l},
        title = "{The data processing pipeline for the MUSE instrument}",
      journal = {\aap},
         year = 2020,
        month = sep,
       volume = {641},
          eid = {A28},
        pages = {A28},
          doi = {10.1051/0004-6361/202037855},
archivePrefix = {arXiv},
       eprint = {2006.08638},
 primaryClass = {astro-ph.IM},
       adsurl = {https://ui.adsabs.harvard.edu/abs/2020A&A...641A..28W}
}

@ARTICLE{2021MNRAS.507.2468S,
       author = {{Scudder}, Jillian M. and {Ellison}, Sara L. and {El Meddah El Idrissi}, Loubna and {Poetrodjojo}, Henry},
        title = "{Conversions between gas-phase metallicities in MaNGA}",
      journal = {\mnras},
         year = 2021,
        month = oct,
       volume = {507},
       number = {2},
        pages = {2468-2487},
          doi = {10.1093/mnras/stab2339},
archivePrefix = {arXiv},
       eprint = {2108.04934},
 primaryClass = {astro-ph.GA},
       adsurl = {https://ui.adsabs.harvard.edu/abs/2021MNRAS.507.2468S}
}

@ARTICLE{2019MNRAS.484.5009E,
       author = {{Erroz-Ferrer}, Santiago and {Carollo}, C. Marcella and {den Brok}, Mark and {Onodera}, Masato and {Brinchmann}, Jarle and {Marino}, Raffaella A. and {Monreal-Ibero}, Ana and {Schaye}, Joop and {Woo}, Joanna and {Cibinel}, Anna and {Debattista}, Victor P. and {Inami}, Hanae and {Maseda}, Michael and {Richard}, Johan and {Tacchella}, Sandro and {Wisotzki}, Lutz},
        title = "{The MUSE Atlas of Disks (MAD): resolving star formation rates and gas metallicities on <100 pc scales{\textdagger}}",
      journal = {\mnras},
         year = 2019,
        month = apr,
       volume = {484},
       number = {4},
        pages = {5009-5027},
          doi = {10.1093/mnras/stz194},
archivePrefix = {arXiv},
       eprint = {1901.04493},
 primaryClass = {astro-ph.GA},
       adsurl = {https://ui.adsabs.harvard.edu/abs/2019MNRAS.484.5009E}
}

@ARTICLE{2020MNRAS.494.1622E,
       author = {{Espinosa-Ponce}, C. and {S{\'a}nchez}, S.~F. and {Morisset}, C. and {Barrera-Ballesteros}, J.~K. and {Galbany}, L. and {Garc{\'\i}a-Benito}, R. and {Lacerda}, E.~A.~D. and {Mast}, D.},
        title = "{H II regions in the CALIFA survey: I. catalogue presentation}",
      journal = {\mnras},
         year = 2020,
        month = may,
       volume = {494},
       number = {2},
        pages = {1622-1646},
          doi = {10.1093/mnras/staa782},
archivePrefix = {arXiv},
       eprint = {2003.07865},
 primaryClass = {astro-ph.GA},
       adsurl = {https://ui.adsabs.harvard.edu/abs/2020MNRAS.494.1622E}
}

@ARTICLE{2022MNRAS.509.1303W,
       author = {{Williams}, Thomas G. and {Kreckel}, Kathryn and {Belfiore}, Francesco and {Groves}, Brent and {Sandstrom}, Karin and {Santoro}, Francesco and {Blanc}, Guillermo A. and {Bigiel}, Frank and {Boquien}, M{\'e}d{\'e}ric and {Chevance}, M{\'e}lanie and {Congiu}, Enrico and {Emsellem}, Eric and {Glover}, Simon C.~O. and {Grasha}, Kathryn and {Klessen}, Ralf S. and {Koch}, Eric and {Kruijssen}, J.~M. Diederik and {Leroy}, Adam K. and {Liu}, Daizhong and {Meidt}, Sharon and {Pan}, Hsi-An and {Querejeta}, Miguel and {Rosolowsky}, Erik and {Saito}, Toshiki and {S{\'a}nchez-Bl{\'a}zquez}, Patricia and {Schinnerer}, Eva and {Schruba}, Andreas and {Watkins}, Elizabeth J.},
        title = "{The 2D metallicity distribution and mixing scales of nearby galaxies}",
      journal = {\mnras},
         year = 2022,
        month = jan,
       volume = {509},
       number = {1},
        pages = {1303-1322},
          doi = {10.1093/mnras/stab3082},
archivePrefix = {arXiv},
       eprint = {2110.10697},
 primaryClass = {astro-ph.GA},
       adsurl = {https://ui.adsabs.harvard.edu/abs/2022MNRAS.509.1303W}
}

@ARTICLE{2014AJ....147..131P,
       author = {{Pilyugin}, L.~S. and {Grebel}, E.~K. and {Kniazev}, A.~Y.},
        title = "{The Abundance Properties of Nearby Late-type Galaxies. I. The Data}",
      journal = {\aj},
         year = 2014,
        month = jun,
       volume = {147},
       number = {6},
          eid = {131},
        pages = {131},
          doi = {10.1088/0004-6256/147/6/131},
archivePrefix = {arXiv},
       eprint = {1403.5461},
 primaryClass = {astro-ph.GA},
       adsurl = {https://ui.adsabs.harvard.edu/abs/2014AJ....147..131P}
}

@ARTICLE{2020MNRAS.492.4149S,
       author = {{S{\'a}nchez-Menguiano}, Laura and {S{\'a}nchez}, Sebasti{\'a}n F. and {P{\'e}rez}, Isabel and {Ruiz-Lara}, Tom{\'a}s and {Galbany}, Llu{\'\i}s and {Anderson}, Joseph P. and {Kuncarayakti}, Hanindyo},
        title = "{Arm-interarm gas abundance variations explored with MUSE: the role of spiral structure in the chemical enrichment of galaxies}",
      journal = {\mnras},
         year = 2020,
        month = mar,
       volume = {492},
       number = {3},
        pages = {4149-4163},
          doi = {10.1093/mnras/staa088},
archivePrefix = {arXiv},
       eprint = {2001.03450},
 primaryClass = {astro-ph.GA},
       adsurl = {https://ui.adsabs.harvard.edu/abs/2020MNRAS.492.4149S}
}

@BOOK{Matteucci2012,
       author = {{Matteucci}, Francesca},
        title = "{Chemical Evolution of Galaxies}",
         year = 2012,
          doi = {10.1007/978-3-642-22491-1},
       adsurl = {https://ui.adsabs.harvard.edu/abs/2012ceg..book.....M}
}

@ARTICLE{Tayal2019,
       author = {{Tayal}, S.~S. and {Zatsarinny}, O. and {Sossah}, A.~M.},
        title = "{Collision and Radiative Rates for Infrared to Extreme Ultraviolet Lines of S III}",
      journal = {\apjs},
         year = 2019,
        month = may,
       volume = {242},
       number = {1},
          eid = {9},
        pages = {9},
          doi = {10.3847/1538-4365/ab17e1},
       adsurl = {https://ui.adsabs.harvard.edu/abs/2019ApJS..242....9T}
}

@BOOK{Osterbrock2006,
       author = {{Osterbrock}, Donald E. and {Ferland}, Gary J.},
        title = "{Astrophysics of gaseous nebulae and active galactic nuclei}",
         year = 2006,
        publisher = {University Science Books},
       adsurl = {https://ui.adsabs.harvard.edu/abs/2006agna.book.....O}
}

@INPROCEEDINGS{Grandmont2012SPIE.8446E..0UG,
       author = {{Grandmont}, F. and {Drissen}, L. and {Mandar}, Julie and {Thibault}, S. and {Baril}, Marc},
        title = "{Final design of SITELLE: a wide-field imaging Fourier transform spectrometer for the Canada-France-Hawaii Telescope}",
    booktitle = {Ground-based and Airborne Instrumentation for Astronomy IV},
         year = 2012,
       editor = {{McLean}, Ian S. and {Ramsay}, Suzanne K. and {Takami}, Hideki},
       series = {Society of Photo-Optical Instrumentation Engineers (SPIE) Conference Series},
       volume = {8446},
        month = sep,
          eid = {84460U},
        pages = {84460U},
          doi = {10.1117/12.926782},
       adsurl = {https://ui.adsabs.harvard.edu/abs/2012SPIE.8446E..0UG}
}

@ARTICLE{Kewley2006,
       author = {{Kewley}, Lisa J. and {Groves}, Brent and {Kauffmann}, Guinevere and {Heckman}, Tim},
        title = "{The host galaxies and classification of active galactic nuclei}",
      journal = {\mnras},
         year = 2006,
        month = nov,
       volume = {372},
       number = {3},
        pages = {961-976},
          doi = {10.1111/j.1365-2966.2006.10859.x},
archivePrefix = {arXiv},
       eprint = {astro-ph/0605681},
 primaryClass = {astro-ph},
       adsurl = {https://ui.adsabs.harvard.edu/abs/2006MNRAS.372..961K}
}

@article{Baldwin1981,
	Adsurl = {http://adsabs.harvard.edu/abs/1981PASP...93....5B},
	Author = {{Baldwin}, J.~A. and {Phillips}, M.~M. and {Terlevich}, R.},
	Doi = {10.1086/130766},
	Journal = {\pasp},
	Month = feb,
	Pages = {5-19},
	Title = {{Classification parameters for the emission-line spectra of extragalactic objects}},
	Volume = 93,
	Year = 1981}

@ARTICLE{RosalesOrtega2026,
       author = {{Rosales-Ortega}, F.~F. and {M{\'e}ndez-Delgado}, J.~E. and {Guerrero-Gonz{\'a}lez}, J.~U. and {Esteban}, C. and {Garc{\'\i}a-Rojas}, J. and {Arellano-C{\'o}rdova}, K.~Z. and {Lugo-Aranda}, A.~Z. and {Esp{\'\i}ndola-Camacho}, O. and {L{\'o}pez-Guti{\'e}rrez}, J.~C. and {Mart{\'\i}nez-Rivero}, L.~E. and {Morisset}, C. and {Orte-Garc{\'\i}a}, M. and {Reyes-Rodr{\'\i}guez}, E. and {Toribio San Cipriano}, L. and {Kreckel}, K. and {Egorov}, O. and {Zinchenko}, I.~A. and {S{\'a}nchez}, S.~F. and {V{\'\i}lchez}, J.~M.},
        title = "{The DESIRED strong-line calibrations: I. New empirical metallicity relations for the local and high-redshift universe}",
      journal = {arXiv e-prints},
         year = 2026,
        month = apr,
          eid = {arXiv:2604.16273},
        pages = {arXiv:2604.16273},
archivePrefix = {arXiv},
       eprint = {2604.16273},
 primaryClass = {astro-ph.GA},
       adsurl = {https://ui.adsabs.harvard.edu/abs/2026arXiv260416273R}
}

@ARTICLE{Stiavelli2025ApJ...981..136S,
       author = {{Stiavelli}, Massimo and {Morishita}, Takahiro and {Chiaberge}, Marco and {Leethochawalit}, Nicha and {Norman}, Colin and {Ricotti}, Massimo and {Roberts-Borsani}, Guido and {Treu}, Tommaso and {Vanzella}, Eros and {Wyse}, Rosemary F.~G. and {Zhang}, Yechi and {Boyett}, Kit},
        title = "{What Can We Learn from the Nitrogen Abundance of High-z Galaxies?}",
      journal = {\apj},
         year = 2025,
        month = mar,
       volume = {981},
       number = {2},
          eid = {136},
        pages = {136},
          doi = {10.3847/1538-4357/adb5f3},
archivePrefix = {arXiv},
       eprint = {2412.06517},
 primaryClass = {astro-ph.GA},
       adsurl = {https://ui.adsabs.harvard.edu/abs/2025ApJ...981..136S}
}

@ARTICLE{Mingozzi2020A&A...636A..42M,
       author = {{Mingozzi}, M. and {Belfiore}, F. and {Cresci}, G. and {Bundy}, K. and {Bershady}, M. and {Bizyaev}, D. and {Blanc}, G. and {Boquien}, M. and {Drory}, N. and {Fu}, H. and {Maiolino}, R. and {Riffel}, R. and {Schaefer}, A. and {Storchi-Bergmann}, T. and {Telles}, E. and {Tremonti}, C. and {Zakamska}, N. and {Zhang}, K.},
        title = "{SDSS IV MaNGA: Metallicity and ionisation parameter in local star-forming galaxies from Bayesian fitting to photoionisation models}",
      journal = {\aap},
         year = 2020,
        month = apr,
       volume = {636},
          eid = {A42},
        pages = {A42},
          doi = {10.1051/0004-6361/201937203},
archivePrefix = {arXiv},
       eprint = {2002.05744},
 primaryClass = {astro-ph.GA},
       adsurl = {https://ui.adsabs.harvard.edu/abs/2020A&A...636A..42M}
}

@ARTICLE{Egorova2026A&A...705A..45E,
       author = {{Egorova}, Evgeniya and {Kreckel}, Kathryn and {Egorov}, Oleg and {Moiseev}, Alexei and {Aragon-Calvo}, Miguel A. and {van de Weygaert}, Rien and {Kotov}, Sergey and {van Gorkom}, Jacqueline},
        title = "{Chemodynamic evidence of pristine gas accretion in the void galaxy VGS 12}",
      journal = {\aap},
         year = 2026,
        month = jan,
       volume = {705},
          eid = {A45},
        pages = {A45},
          doi = {10.1051/0004-6361/202557058},
archivePrefix = {arXiv},
       eprint = {2511.00235},
 primaryClass = {astro-ph.GA},
       adsurl = {https://ui.adsabs.harvard.edu/abs/2026A&A...705A..45E}
}

@ARTICLE{Ho2019MNRAS.485.3569H,
       author = {{Ho}, I.-Ting},
        title = "{A machine learning artificial neural network calibration of the strong-line oxygen abundance}",
      journal = {\mnras},
         year = 2019,
        month = may,
       volume = {485},
       number = {3},
        pages = {3569-3579},
          doi = {10.1093/mnras/stz649},
archivePrefix = {arXiv},
       eprint = {1903.01506},
 primaryClass = {astro-ph.GA},
       adsurl = {https://ui.adsabs.harvard.edu/abs/2019MNRAS.485.3569H}
}

@ARTICLE{Perez-Montero2014MNRAS.441.2663P,
       author = {{P{\'e}rez-Montero}, E.},
        title = "{Deriving model-based T$_{e}$-consistent chemical abundances in ionized gaseous nebulae}",
      journal = {\mnras},
         year = 2014,
        month = jul,
       volume = {441},
       number = {3},
        pages = {2663-2675},
          doi = {10.1093/mnras/stu753},
archivePrefix = {arXiv},
       eprint = {1404.3936},
 primaryClass = {astro-ph.GA},
       adsurl = {https://ui.adsabs.harvard.edu/abs/2014MNRAS.441.2663P}
}

@ARTICLE{Grasha2022ApJ...929..118G,
       author = {{Grasha}, K. and {Chen}, Q.~H. and {Battisti}, A.~J. and {Acharyya}, A. and {Ridolfo}, S. and {Poehler}, E. and {Mably}, S. and {Verma}, A.~A. and {Hayward}, K.~L. and {Kharbanda}, A. and {Poetrodjojo}, H. and {Seibert}, M. and {Rich}, J.~A. and {Madore}, B.~F. and {Kewley}, L.~J.},
        title = "{Metallicity, Ionization Parameter, and Pressure Variations of H II Regions in the TYPHOON Spiral Galaxies: NGC 1566, NGC 2835, NGC 3521, NGC 5068, NGC 5236, and NGC 7793}",
      journal = {\apj},
         year = 2022,
        month = apr,
       volume = {929},
       number = {2},
          eid = {118},
        pages = {118},
          doi = {10.3847/1538-4357/ac5ab2},
archivePrefix = {arXiv},
       eprint = {2203.02522},
 primaryClass = {astro-ph.GA},
       adsurl = {https://ui.adsabs.harvard.edu/abs/2022ApJ...929..118G}
}

@ARTICLE{Rousseau-Nepton2019MNRAS.489.5530R,
       author = {{Rousseau-Nepton}, L. and {Martin}, R.~P. and {Robert}, C. and {Drissen}, L. and {Amram}, P. and {Prunet}, S. and {Martin}, T. and {Moumen}, I. and {Adamo}, A. and {Alarie}, A. and {Barmby}, P. and {Boselli}, A. and {Bresolin}, F. and {Bureau}, M. and {Chemin}, L. and {Fernandes}, R.~C. and {Combes}, F. and {Crowder}, C. and {Della Bruna}, L. and {Duarte Puertas}, S. and {Egusa}, F. and {Epinat}, B. and {Ksoll}, V.~F. and {Girard}, M. and {G{\'o}mez Llanos}, V. and {Gouliermis}, D. and {Grasha}, K. and {Higgs}, C. and {Hlavacek-Larrondo}, J. and {Ho}, I.-T. and {Iglesias-P{\'a}ramo}, J. and {Joncas}, G. and {Kam}, Z.~S. and {Karera}, P. and {Kennicutt}, R.~C. and {Klessen}, R.~S. and {Lianou}, S. and {Liu}, L. and {Liu}, Q. and {de Amorim}, A. Luiz and {Lyman}, J.~D. and {Martel}, H. and {Mazzilli-Ciraulo}, B. and {McLeod}, A.~F. and {Melchior}, A.-L. and {Millan}, I. and {Moll{\'a}}, M. and {Momose}, R. and {Morisset}, C. and {Pan}, H.-A. and {Pati}, A.~K. and {Pellerin}, A. and {Pellegrini}, E. and {P{\'e}rez}, I. and {Petric}, A. and {Plana}, H. and {Rahner}, D. and {Ruiz Lara}, T. and {S{\'a}nchez-Menguiano}, L. and {Spekkens}, K. and {Stasi{\'n}ska}, G. and {Takamiya}, M. and {Vale Asari}, N. and {V{\'\i}lchez}, J.~M.},
        title = "{SIGNALS: I. Survey description}",
      journal = {\mnras},
         year = 2019,
        month = nov,
       volume = {489},
       number = {4},
        pages = {5530-5546},
          doi = {10.1093/mnras/stz2455},
archivePrefix = {arXiv},
       eprint = {1908.09017},
 primaryClass = {astro-ph.GA},
       adsurl = {https://ui.adsabs.harvard.edu/abs/2019MNRAS.489.5530R}
}

@ARTICLE{PerezMontero2009MNRAS.398..949P,
       author = {{P{\'e}rez-Montero}, Enrique and {Contini}, Thierry},
        title = "{The impact of the nitrogen-to-oxygen ratio on ionized nebula diagnostics based on [NII] emission lines}",
      journal = {\mnras},
         year = 2009,
        month = sep,
       volume = {398},
       number = {2},
        pages = {949-960},
          doi = {10.1111/j.1365-2966.2009.15145.x},
archivePrefix = {arXiv},
       eprint = {0905.4621},
 primaryClass = {astro-ph.CO},
       adsurl = {https://ui.adsabs.harvard.edu/abs/2009MNRAS.398..949P}
}

@ARTICLE{Blanc2015ApJ...798...99B,
       author = {{Blanc}, Guillermo A. and {Kewley}, Lisa and {Vogt}, Fr{\'e}d{\'e}ric P.~A. and {Dopita}, Michael A.},
        title = "{IZI: Inferring the Gas Phase Metallicity (Z) and Ionization Parameter (q) of Ionized Nebulae Using Bayesian Statistics}",
      journal = {\apj},
         year = 2015,
        month = jan,
       volume = {798},
       number = {2},
          eid = {99},
        pages = {99},
          doi = {10.1088/0004-637X/798/2/99},
archivePrefix = {arXiv},
       eprint = {1410.8146},
 primaryClass = {astro-ph.GA},
       adsurl = {https://ui.adsabs.harvard.edu/abs/2015ApJ...798...99B}
}

@ARTICLE{Boissier1999MNRAS.307..857B,
       author = {{Boissier}, S. and {Prantzos}, N.},
        title = "{Chemo-spectrophotometric evolution of spiral galaxies - I. The model and the Milky Way}",
      journal = {\mnras},
         year = 1999,
        month = aug,
       volume = {307},
       number = {4},
        pages = {857-876},
          doi = {10.1046/j.1365-8711.1999.02699.x},
archivePrefix = {arXiv},
       eprint = {astro-ph/9902148},
 primaryClass = {astro-ph},
       adsurl = {https://ui.adsabs.harvard.edu/abs/1999MNRAS.307..857B}
}

@ARTICLE{Poetrodjojo2018MNRAS.479.5235P,
       author = {{Poetrodjojo}, Henry and {Groves}, Brent and {Kewley}, Lisa J. and {Medling}, Anne M. and {Sweet}, Sarah M. and {van de Sande}, Jesse and {Sanchez}, Sebastian F. and {Bland-Hawthorn}, Joss and {Brough}, Sarah and {Bryant}, Julia J. and {Cortese}, Luca and {Croom}, Scott M. and {L{\'o}pez-S{\'a}nchez}, {\'A}ngel R. and {Richards}, Samuel N. and {Zafar}, Tayyaba and {Lawrence}, Jon S. and {Lorente}, Nuria P.~F. and {Owers}, Matt S. and {Scott}, Nicholas},
        title = "{The SAMI Galaxy Survey: Spatially resolved metallicity and ionization mapping}",
      journal = {\mnras},
         year = 2018,
        month = oct,
       volume = {479},
       number = {4},
        pages = {5235-5265},
          doi = {10.1093/mnras/sty1782},
archivePrefix = {arXiv},
       eprint = {1807.01522},
 primaryClass = {astro-ph.GA},
       adsurl = {https://ui.adsabs.harvard.edu/abs/2018MNRAS.479.5235P}
}

@ARTICLE{Belfiore2017MNRAS.469..151B,
       author = {{Belfiore}, Francesco and {Maiolino}, Roberto and {Tremonti}, Christy and {S{\'a}nchez}, Sebastian F. and {Bundy}, Kevin and {Bershady}, Matthew and {Westfall}, Kyle and {Lin}, Lihwai and {Drory}, Niv and {Boquien}, M{\'e}d{\'e}ric and {Thomas}, Daniel and {Brinkmann}, Jonathan},
        title = "{SDSS IV MaNGA - metallicity and nitrogen abundance gradients in local galaxies}",
      journal = {\mnras},
         year = 2017,
        month = jul,
       volume = {469},
       number = {1},
        pages = {151-170},
          doi = {10.1093/mnras/stx789},
archivePrefix = {arXiv},
       eprint = {1703.03813},
 primaryClass = {astro-ph.GA},
       adsurl = {https://ui.adsabs.harvard.edu/abs/2017MNRAS.469..151B}
}

@ARTICLE{Kaplan2016MNRAS.462.1642K,
       author = {{Kaplan}, Kyle F. and {Jogee}, Shardha and {Kewley}, Lisa and {Blanc}, Guillermo A. and {Weinzirl}, Tim and {Song}, Mimi and {Drory}, Niv and {Luo}, Rongxin and {van den Bosch}, Remco C.~E.},
        title = "{The VIRUS-P Exploration of Nearby Galaxies (VENGA): spatially resolved gas-phase metallicity distributions in barred and unbarred spirals}",
      journal = {\mnras},
         year = 2016,
        month = oct,
       volume = {462},
       number = {2},
        pages = {1642-1682},
          doi = {10.1093/mnras/stw1422},
archivePrefix = {arXiv},
       eprint = {1606.05713},
 primaryClass = {astro-ph.GA},
       adsurl = {https://ui.adsabs.harvard.edu/abs/2016MNRAS.462.1642K}
}

@ARTICLE{Sanchez2014A&A...563A..49S,
       author = {{S{\'a}nchez}, S.~F. and {Rosales-Ortega}, F.~F. and {Iglesias-P{\'a}ramo}, J. and {Moll{\'a}}, M. and {Barrera-Ballesteros}, J. and {Marino}, R.~A. and {P{\'e}rez}, E. and {S{\'a}nchez-Blazquez}, P. and {Gonz{\'a}lez Delgado}, R. and {Cid Fernandes}, R. and {de Lorenzo-C{\'a}ceres}, A. and {Mendez-Abreu}, J. and {Galbany}, L. and {Falcon-Barroso}, J. and {Miralles-Caballero}, D. and {Husemann}, B. and {Garc{\'\i}a-Benito}, R. and {Mast}, D. and {Walcher}, C.~J. and {Gil de Paz}, A. and {Garc{\'\i}a-Lorenzo}, B. and {Jungwiert}, B. and {V{\'\i}lchez}, J.~M. and {J{\'\i}lkov{\'a}}, Lucie and {Lyubenova}, M. and {Cortijo-Ferrero}, C. and {D{\'\i}az}, A.~I. and {Wisotzki}, L. and {M{\'a}rquez}, I. and {Bland-Hawthorn}, J. and {Ellis}, S. and {van de Ven}, G. and {Jahnke}, K. and {Papaderos}, P. and {Gomes}, J.~M. and {Mendoza}, M.~A. and {L{\'o}pez-S{\'a}nchez}, {\'A}. R.},
        title = "{A characteristic oxygen abundance gradient in galaxy disks unveiled with CALIFA}",
      journal = {\aap},
         year = 2014,
        month = mar,
       volume = {563},
          eid = {A49},
        pages = {A49},
          doi = {10.1051/0004-6361/201322343},
archivePrefix = {arXiv},
       eprint = {1311.7052},
 primaryClass = {astro-ph.CO},
       adsurl = {https://ui.adsabs.harvard.edu/abs/2014A&A...563A..49S}
}

@ARTICLE{Zaritsky1994ApJ...420...87Z,
       author = {{Zaritsky}, Dennis and {Kennicutt}, Jr., Robert C. and {Huchra}, John P.},
        title = "{H II Regions and the Abundance Properties of Spiral Galaxies}",
      journal = {\apj},
         year = 1994,
        month = jan,
       volume = {420},
        pages = {87},
          doi = {10.1086/173544},
       adsurl = {https://ui.adsabs.harvard.edu/abs/1994ApJ...420...87Z}
}

@ARTICLE{Kennicutt1996ApJ...456..504K,
       author = {{Kennicutt}, Jr., Robert C. and {Garnett}, Donald R.},
        title = "{The Composition Gradient in M101 Revisited. I. H II Region Spectra and Excitation Properties}",
      journal = {\apj},
         year = 1996,
        month = jan,
       volume = {456},
        pages = {504},
          doi = {10.1086/176675},
       adsurl = {https://ui.adsabs.harvard.edu/abs/1996ApJ...456..504K}
}

@ARTICLE{Bresolin2025MNRAS.539..755B,
       author = {{Bresolin}, Fabio and {Fern{\'a}ndez-Arenas}, David and {Rousseau-Nepton}, Laurie and {Garner}, III, Ray and {Zurita}, Almudena and {Robert}, Carmelle and {Drissen}, Laurent and {Martin}, Ren{\'e} Pierre and {Amram}, Philippe and {Duarte Puertas}, Salvador and {Savard}, Gabriel and {Vicens}, S{\'e}bastien and {Posternak}, Mykola},
        title = "{SIGNALS on the mixing of oxygen and nitrogen in the spiral galaxy NGC 6946}",
      journal = {\mnras},
         year = 2025,
        month = may,
       volume = {539},
       number = {2},
        pages = {755-770},
          doi = {10.1093/mnras/staf510},
archivePrefix = {arXiv},
       eprint = {2502.00649},
 primaryClass = {astro-ph.GA},
       adsurl = {https://ui.adsabs.harvard.edu/abs/2025MNRAS.539..755B}
}

@ARTICLE{Hwang2019ApJ...872..144H,
       author = {{Hwang}, Hsiang-Chih and {Barrera-Ballesteros}, Jorge K. and {Heckman}, Timothy M. and {Rowlands}, Kate and {Lin}, Lihwai and {Rodriguez-Gomez}, Vicente and {Pan}, Hsi-An and {Hsieh}, Bau-Ching and {S{\'a}nchez}, Sebastian and {Bizyaev}, Dmitry and {S{\'a}nchez Almeida}, Jorge and {Thilker}, David A. and {Lotz}, Jennifer M. and {Jones}, Amy and {Nair}, Preethi and {Andrews}, Brett H. and {Drory}, Niv},
        title = "{Anomalously Low-metallicity Regions in MaNGA Star-forming Galaxies: Accretion Caught in Action?}",
      journal = {\apj},
         year = 2019,
        month = feb,
       volume = {872},
       number = {2},
          eid = {144},
        pages = {144},
          doi = {10.3847/1538-4357/aaf7a3},
archivePrefix = {arXiv},
       eprint = {1812.04614},
 primaryClass = {astro-ph.GA},
       adsurl = {https://ui.adsabs.harvard.edu/abs/2019ApJ...872..144H}
}

@ARTICLE{Garner2025arXiv251003144G,
       author = {{Garner}, III, Ray and {Kennicutt}, Jr, Robert C. and {Drissen}, Laurent and {Robert}, Carmelle and {Rousseau-Nepton}, Laurie and {Morisset}, Christophe and {Amram}, Philippe and {Martin}, R. Pierre and {Jarvis}, Emma},
        title = "{SIGNALS of Giant HII Regions: A Spatially Resolved Analysis of NGC 604}",
      journal = {arXiv e-prints},
         year = 2025,
        month = oct,
          eid = {arXiv:2510.03144},
        pages = {arXiv:2510.03144},
          doi = {10.48550/arXiv.2510.03144},
archivePrefix = {arXiv},
       eprint = {2510.03144},
 primaryClass = {astro-ph.GA},
       adsurl = {https://ui.adsabs.harvard.edu/abs/2025arXiv251003144G}
}

@ARTICLE{Maiolino2019A&ARv..27....3M,
       author = {{Maiolino}, R. and {Mannucci}, F.},
        title = "{De re metallica: the cosmic chemical evolution of galaxies}",
      journal = {\aapr},
         year = 2019,
        month = feb,
       volume = {27},
       number = {1},
          eid = {3},
        pages = {3},
          doi = {10.1007/s00159-018-0112-2},
archivePrefix = {arXiv},
       eprint = {1811.09642},
 primaryClass = {astro-ph.GA},
       adsurl = {https://ui.adsabs.harvard.edu/abs/2019A&ARv..27....3M}
}

@ARTICLE{Kreckel2020MNRAS.499..193K,
       author = {{Kreckel}, Kathryn and {Ho}, I.-Ting and {Blanc}, Guillermo A. and {Glover}, Simon C.~O. and {Groves}, Brent and {Rosolowsky}, Erik and {Bigiel}, Frank and {Boqu{\'\i}en}, M{\'e}d{\'e}ric and {Chevance}, M{\'e}lanie and {Dale}, Daniel A. and {Deger}, Sinan and {Emsellem}, Eric and {Grasha}, Kathryn and {Kim}, Jenny J. and {Klessen}, Ralf S. and {Kruijssen}, J.~M. Diederik and {Lee}, Janice C. and {Leroy}, Adam K. and {Liu}, Daizhong and {McElroy}, Rebecca and {Meidt}, Sharon E. and {Pessa}, Ismael and {Sanchez-Blazquez}, Patricia and {Sandstrom}, Karin and {Santoro}, Francesco and {Scheuermann}, Fabian and {Schinnerer}, Eva and {Schruba}, Andreas and {Utomo}, Dyas and {Watkins}, Elizabeth J. and {Williams}, Thomas G.},
        title = "{Measuring the mixing scale of the ISM within nearby spiral galaxies}",
      journal = {\mnras},
         year = 2020,
        month = nov,
       volume = {499},
       number = {1},
        pages = {193-209},
          doi = {10.1093/mnras/staa2743},
archivePrefix = {arXiv},
       eprint = {2009.02342},
 primaryClass = {astro-ph.GA},
       adsurl = {https://ui.adsabs.harvard.edu/abs/2020MNRAS.499..193K}
}

@ARTICLE{Berg2020ApJ...893...96B,
       author = {{Berg}, Danielle A. and {Pogge}, Richard W. and {Skillman}, Evan D. and {Croxall}, Kevin V. and {Moustakas}, John and {Rogers}, Noah S.~J. and {Sun}, Jiayi},
        title = "{CHAOS IV: Gas-phase Abundance Trends from the First Four CHAOS Galaxies}",
      journal = {\apj},
         year = 2020,
        month = apr,
       volume = {893},
       number = {2},
          eid = {96},
        pages = {96},
          doi = {10.3847/1538-4357/ab7eab},
archivePrefix = {arXiv},
       eprint = {2001.05002},
 primaryClass = {astro-ph.GA},
       adsurl = {https://ui.adsabs.harvard.edu/abs/2020ApJ...893...96B}
}

@ARTICLE{Li2024A&A...690A.161L,
       author = {{Li}, Jing and {Kreckel}, K. and {Sarbadhicary}, S. and {Egorov}, Oleg V. and {Groves}, B. and {Long}, K.~S. and {Congiu}, Enrico and {Belfiore}, Francesco and {Glover}, Simon C.~O. and {Barnes}, Ashley T. and {Bigiel}, Frank and {Blanc}, Guillermo A. and {Grasha}, Kathryn and {Klessen}, Ralf S. and {Leroy}, Adam and {Lopez}, Laura A. and {M{\'e}ndez-Delgado}, J. Eduardo and {Neumann}, Justus and {Schinnerer}, Eva and {Williams}, Thomas G.},
        title = "{Discovery of  2200 new supernova remnants in 19 nearby star-forming galaxies with MUSE spectroscopy}",
      journal = {\aap},
         year = 2024,
        month = oct,
       volume = {690},
          eid = {A161},
        pages = {A161},
          doi = {10.1051/0004-6361/202450730},
archivePrefix = {arXiv},
       eprint = {2405.08974},
 primaryClass = {astro-ph.GA},
       adsurl = {https://ui.adsabs.harvard.edu/abs/2024A&A...690A.161L}
}

@ARTICLE{Kreckel2025A&A...703A..42K,
       author = {{Kreckel}, K. and {Rickards Vaught}, R.~J. and {Egorov}, O.~V. and {M{\'e}ndez-Delgado}, J.~E. and {Belfiore}, F. and {Brazzini}, M. and {Egorova}, E. and {Congiu}, E. and {Dale}, D.~A. and {Dlamini}, S. and {Glover}, S.~C.~O. and {Grasha}, K. and {Klessen}, R.~S. and {Liang}, F.-H. and {Pan}, H.-A. and {S{\'a}nchez-Bl{\'a}zquez}, P. and {Williams}, T.~G.},
        title = "{Temperature-based radial metallicity gradients in nearby galaxies}",
      journal = {\aap},
         year = 2025,
        month = nov,
       volume = {703},
          eid = {A42},
        pages = {A42},
          doi = {10.1051/0004-6361/202556017},
archivePrefix = {arXiv},
       eprint = {2507.20744},
 primaryClass = {astro-ph.GA},
       adsurl = {https://ui.adsabs.harvard.edu/abs/2025A&A...703A..42K}
}

@ARTICLE{RickardsVaught2024ApJ...966..130R,
       author = {{Rickards Vaught}, Ryan J. and {Sandstrom}, Karin M. and {Belfiore}, Francesco and {Kreckel}, Kathryn and {M{\'e}ndez-Delgado}, J. Eduardo and {Emsellem}, Eric and {Groves}, Brent and {Blanc}, Guillermo A. and {Dale}, Daniel A. and {Egorov}, Oleg V. and {Glover}, Simon C.~O. and {Grasha}, Kathryn and {Klessen}, Ralf S. and {Neumann}, Justus and {Williams}, Thomas G.},
        title = "{Investigating the Drivers of Electron Temperature Variations in H II Regions with Keck-KCWI and VLT-MUSE}",
      journal = {\apj},
         year = 2024,
        month = may,
       volume = {966},
       number = {1},
          eid = {130},
        pages = {130},
          doi = {10.3847/1538-4357/ad303c},
archivePrefix = {arXiv},
       eprint = {2309.17440},
 primaryClass = {astro-ph.GA},
       adsurl = {https://ui.adsabs.harvard.edu/abs/2024ApJ...966..130R}
}

@ARTICLE{Brazzini2024A&A...691A.173B,
       author = {{Brazzini}, Matilde and {Belfiore}, Francesco and {Ginolfi}, Michele and {Groves}, Brent and {Kreckel}, Kathryn and {Rickards Vaught}, Ryan J. and {Baron}, Dalya and {Bigiel}, Frank and {Blanc}, Guillermo A. and {Dale}, Daniel A. and {Grasha}, Kathryn and {Habjan}, Eric and {Klessen}, Ralf S. and {M{\'e}ndez-Delgado}, Jose Eduardo and {Sandstrom}, Karin and {Williams}, Thomas G.},
        title = "{Metallicity calibrations based on auroral lines from PHANGS{\textendash}MUSE data}",
      journal = {\aap},
         year = 2024,
        month = nov,
       volume = {691},
          eid = {A173},
        pages = {A173},
          doi = {10.1051/0004-6361/202451007},
archivePrefix = {arXiv},
       eprint = {2410.00106},
 primaryClass = {astro-ph.GA},
       adsurl = {https://ui.adsabs.harvard.edu/abs/2024A&A...691A.173B}
}

@ARTICLE{Emsellem2022,
       author = {{Emsellem}, Eric and {Schinnerer}, Eva and {Santoro}, Francesco and {Belfiore}, Francesco and {Pessa}, Ismael and {McElroy}, Rebecca and {Blanc}, Guillermo A. and {Congiu}, Enrico and {Groves}, Brent and {Ho}, I. -Ting and {Kreckel}, Kathryn and {Razza}, Alessandro and {Sanchez-Blazquez}, Patricia and {Egorov}, Oleg and {Faesi}, Chris and {Klessen}, Ralf S. and {Leroy}, Adam K. and {Meidt}, Sharon and {Querejeta}, Miguel and {Rosolowsky}, Erik and {Scheuermann}, Fabian and {Anand}, Gagandeep S. and {Barnes}, Ashley T. and {Be{\v{s}}li{\'c}}, Ivana and {Bigiel}, Frank and {Boquien}, M{\'e}d{\'e}ric and {Cao}, Yixian and {Chevance}, M{\'e}lanie and {Dale}, Daniel A. and {Eibensteiner}, Cosima and {Glover}, Simon C.~O. and {Grasha}, Kathryn and {Henshaw}, Jonathan D. and {Hughes}, Annie and {Koch}, Eric W. and {Kruijssen}, J.~M. Diederik and {Lee}, Janice and {Liu}, Daizhong and {Pan}, Hsi-An and {Pety}, J{\'e}r{\^o}me and {Saito}, Toshiki and {Sandstrom}, Karin M. and {Schruba}, Andreas and {Sun}, Jiayi and {Thilker}, David A. and {Usero}, Antonio and {Watkins}, Elizabeth J. and {Williams}, Thomas G.},
        title = "{The PHANGS-MUSE survey. Probing the chemo-dynamical evolution of disc galaxies}",
      journal = {\aap},
         year = 2022,
        month = mar,
       volume = {659},
          eid = {A191},
        pages = {A191},
          doi = {10.1051/0004-6361/202141727},
archivePrefix = {arXiv},
       eprint = {2110.03708},
 primaryClass = {astro-ph.GA},
       adsurl = {https://ui.adsabs.harvard.edu/abs/2022A&A...659A.191E}
}

@article{Pilyugin2016,
   title={New calibrations for abundance determinations in HII regions},
   volume={457},
   ISSN={1365-2966},
   url={http://dx.doi.org/10.1093/mnras/stw238},
   DOI={10.1093/mnras/stw238},
   number={4},
   journal={Monthly Notices of the Royal Astronomical Society},
   publisher={Oxford University Press (OUP)},
   author={Pilyugin, L. S. and Grebel, E. K.},
   year={2016},
   month=jan, pages={3678–3692} }

@ARTICLE{Kauffmann2003,
       author = {{Kauffmann}, Guinevere and {Heckman}, Timothy M. and {Tremonti}, Christy and {Brinchmann}, Jarle and {Charlot}, St{\'e}phane and {White}, Simon D.~M. and {Ridgway}, Susan E. and {Brinkmann}, Jon and {Fukugita}, Masataka and {Hall}, Patrick B. and {Ivezi{\'c}}, {\v{Z}}eljko and {Richards}, Gordon T. and {Schneider}, Donald P.},
        title = "{The host galaxies of active galactic nuclei}",
      journal = {\mnras},
         year = 2003,
        month = dec,
       volume = {346},
       number = {4},
        pages = {1055-1077},
          doi = {10.1111/j.1365-2966.2003.07154.x},
archivePrefix = {arXiv},
       eprint = {astro-ph/0304239},
 primaryClass = {astro-ph},
       adsurl = {https://ui.adsabs.harvard.edu/abs/2003MNRAS.346.1055K}
}

@Misc{ESOMUSE,
  Author =    {{Bacon}, R. and {Glindemann}, A. and {Vernet}, J. and {Manescau}, A.},
  title =     "{A Multi Unit Spectroscopic Explorer - MUSE}",
  year = "2005",
  url = "https://www.eso.org/sci/facilities/develop/instruments/muse.html"
   }

@INPROCEEDINGS{ORBS_ORCS,
       author = {{Martin}, T. and {Drissen}, L. and {Joncas}, G.},
        title = "{ORBS, ORCS, OACS, a Software Suite for Data Reduction and Analysis of the Hyperspectral Imagers SITELLE and SpIOMM}",
    booktitle = {Astronomical Data Analysis Software an Systems XXIV (ADASS XXIV)},
         year = 2015,
       editor = {{Taylor}, A.~R. and {Rosolowsky}, E.},
       series = {Astronomical Society of the Pacific Conference Series},
       volume = {495},
        month = sep,
        pages = {327},
       adsurl = {https://ui.adsabs.harvard.edu/abs/2015ASPC..495..327M}
}

@ARTICLE{Kewley2002,
       author = {{Kewley}, L.~J. and {Dopita}, M.~A.},
        title = "{Using Strong Lines to Estimate Abundances in Extragalactic H II Regions and Starburst Galaxies}",
      journal = {\apjs},
         year = 2002,
        month = sep,
       volume = {142},
       number = {1},
        pages = {35-52},
          doi = {10.1086/341326},
archivePrefix = {arXiv},
       eprint = {astro-ph/0206495},
 primaryClass = {astro-ph},
       adsurl = {https://ui.adsabs.harvard.edu/abs/2002ApJS..142...35K}
}

@ARTICLE{ODonnnell1994,
       author = {{O'Donnell}, James E.},
        title = "{R v-dependent Optical and Near-Ultraviolet Extinction}",
      journal = {\apj},
         year = 1994,
        month = feb,
       volume = {422},
        pages = {158},
          doi = {10.1086/173713},
       adsurl = {https://ui.adsabs.harvard.edu/abs/1994ApJ...422..158O}
}

@Misc{pyneb,
  Author =    {{Luridiana}, V. and {Morisset}, C. and {Shaw}, R. A},
  title =     "{PyNeb Reference Manual}",
  year = "2022",
  url = "https://pypi.org/project/PyNeb/"
   }

@ARTICLE{Luridiana2015,
   title={PyNeb: a new tool for analyzing emission lines: I. Code description and validation of results},
   volume={573},
   ISSN={1432-0746},
   url={http://dx.doi.org/10.1051/0004-6361/201323152},
   DOI={10.1051/0004-6361/201323152},
   journal={Astronomy &amp; Astrophysics},
   publisher={EDP Sciences},
   author={Luridiana, V. and Morisset, C. and Shaw, R. A.},
   year={2014},
   month=dec, pages={A42} }

@ARTICLE{Groves2023,
   title={The PHANGS–MUSE nebular catalogue},
   volume={520},
   ISSN={1365-2966},
   url={http://dx.doi.org/10.1093/mnras/stad114},
   DOI={10.1093/mnras/stad114},
   number={4},
   journal={Monthly Notices of the Royal Astronomical Society},
   publisher={Oxford University Press (OUP)},
   author={Groves, B and Kreckel, K and Santoro, F and Belfiore, F and Zavodnik, E and Congiu, E and Egorov, O V and Emsellem, E and Grasha, K and Leroy, A and Scheuermann, F and Schinnerer, E and Watkins, E J and Barnes, A T and Bigiel, F and Dale, D A and Glover, S C O and Pessa, I and Sanchez-Blazquez, P and Williams, T G},
   year={2023},
   month=jan, pages={4902–4952} }

@ARTICLE{Thilker2000,
   title={HII[CLC]phot[/CLC]: Automated Photometry of H [CSC]ii[/CSC] Regions Applied to M51},
   volume={120},
   ISSN={0004-6256},
   url={http://dx.doi.org/10.1086/316852},
   DOI={10.1086/316852},
   number={6},
   journal={The Astronomical Journal},
   publisher={American Astronomical Society},
   author={Thilker, David A. and Braun, Robert and Walterbos, René A. M.},
   year={2000},
   month=dec, pages={3070–3087} }

@ARTICLE{Rousseau-Nepton2018,
       author = {{Rousseau-Nepton}, L. and {Robert}, C. and {Martin}, R.~P. and {Drissen}, L. and {Martin}, T.},
        title = "{NGC628 with SITELLE: I. Imaging spectroscopy of 4285 H II region candidates}",
      journal = {\mnras},
         year = 2018,
        month = feb,
       volume = {477},
       number = {3},
        pages = {4152-4186},
          doi = {10.1093/mnras/sty477},
archivePrefix = {arXiv},
       eprint = {1704.05121},
 primaryClass = {astro-ph.GA},
       adsurl = {https://ui.adsabs.harvard.edu/abs/2018MNRAS.477.4152R}
}

@Misc{photutils,
author       = {Larry Bradley and
                Brigitta Sip{\H o}cz and
                Thomas Robitaille and
                Erik Tollerud and
                Z\`e Vin{\'{\i}}cius and
                Christoph Deil and
                Kyle Barbary and
                Tom J Wilson and
                Ivo Busko and
                Axel Donath and
                Hans Moritz G{\"u}nther and
                Mihai Cara and
                P. L. Lim and
                Sebastian Me{\ss}linger and
                Simon Conseil and
                Azalee Bostroem and
                Michael Droettboom and
                E. M. Bray and
                Lars Andersen Bratholm and
                Geert Barentsen and
                Matt Craig and
                Shivangee Rathi and
                Sergio Pascual and
                Gabriel Perren and
                Iskren Y. Georgiev and
                Miguel de Val-Borro and
                Wolfgang Kerzendorf and
                Yoonsoo P. Bach and
                Bruno Quint and
                Harrison Souchereau},
title        = {astropy/photutils: 1.8.0},
month        = may,
year         = 2023,
publisher    = {Zenodo},
version      = {1.8.0},
doi          = {10.5281/zenodo.7946442},
url          = {https://doi.org/10.5281/zenodo.7946442}
}

@ARTICLE{Martin2016,
       author = {{Martin}, Thomas B. and {Prunet}, Simon and {Drissen}, Laurent},
        title = "{Optimal fitting of Gaussian-apodized or under-resolved emission lines in Fourier transform spectra providing new insights on the velocity structure of NGC 6720}",
      journal = {\mnras},
         year = 2016,
        month = dec,
       volume = {463},
       number = {4},
        pages = {4223-4238},
          doi = {10.1093/mnras/stw2315},
archivePrefix = {arXiv},
       eprint = {1608.05854},
 primaryClass = {astro-ph.GA},
       adsurl = {https://ui.adsabs.harvard.edu/abs/2016MNRAS.463.4223M}
}

@ARTICLE{Makarov2014,
       author = {{Makarov}, Dmitry and {Prugniel}, Philippe and {Terekhova}, Nataliya and {Courtois}, H{\'e}l{\`e}ne and {Vauglin}, Isabelle},
        title = "{HyperLEDA. III. The catalogue of extragalactic distances}",
      journal = {\aap},
         year = 2014,
        month = oct,
       volume = {570},
          eid = {A13},
        pages = {A13},
          doi = {10.1051/0004-6361/201423496},
archivePrefix = {arXiv},
       eprint = {1408.3476},
 primaryClass = {astro-ph.GA},
       adsurl = {https://ui.adsabs.harvard.edu/abs/2014A&A...570A..13M}
}

@ARTICLE{Anand2021,
       author = {{Anand}, Gagandeep S. and {Lee}, Janice C. and {Van Dyk}, Schuyler D. and {Leroy}, Adam K. and {Rosolowsky}, Erik and {Schinnerer}, Eva and {Larson}, Kirsten and {Kourkchi}, Ehsan and {Kreckel}, Kathryn and {Scheuermann}, Fabian and {Rizzi}, Luca and {Thilker}, David and {Tully}, R. Brent and {Bigiel}, Frank and {Blanc}, Guillermo A. and {Boquien}, M{\'e}d{\'e}ric and {Chandar}, Rupali and {Dale}, Daniel and {Emsellem}, Eric and {Deger}, Sinan and {Glover}, Simon C.~O. and {Grasha}, Kathryn and {Groves}, Brent and {S. Klessen}, Ralf and {Kruijssen}, J.~M. Diederik and {Querejeta}, Miguel and {S{\'a}nchez-Bl{\'a}zquez}, Patricia and {Schruba}, Andreas and {Turner}, Jordan and {Ubeda}, Leonardo and {Williams}, Thomas G. and {Whitmore}, Brad},
        title = "{Distances to PHANGS galaxies: New tip of the red giant branch measurements and adopted distances}",
      journal = {\mnras},
         year = 2021,
        month = mar,
       volume = {501},
       number = {3},
        pages = {3621-3639},
          doi = {10.1093/mnras/staa3668},
archivePrefix = {arXiv},
       eprint = {2012.00757},
 primaryClass = {astro-ph.GA},
       adsurl = {https://ui.adsabs.harvard.edu/abs/2021MNRAS.501.3621A}
}

@ARTICLE{Leroy2021,
   title={PHANGS–ALMA: Arcsecond CO(2–1) Imaging of Nearby Star-forming Galaxies},
   volume={257},
   ISSN={1538-4365},
   url={http://dx.doi.org/10.3847/1538-4365/ac17f3},
   DOI={10.3847/1538-4365/ac17f3},
   number={2},
   journal={The Astrophysical Journal Supplement Series},
   publisher={American Astronomical Society},
   author={Leroy, Adam K. and Schinnerer, Eva and Hughes, Annie and Rosolowsky, Erik and Pety, Jérôme and Schruba, Andreas and Usero, Antonio and Blanc, Guillermo A. and Chevance, Mélanie and Emsellem, Eric and Faesi, Christopher M. and Herrera, Cinthya N. and Liu, Daizhong and Meidt, Sharon E. and Querejeta, Miguel and Saito, Toshiki and Sandstrom, Karin M. and Sun 孙, Jiayi 嘉 懿 and Williams, Thomas G. and Anand, Gagandeep S. and Barnes, Ashley T. and Behrens, Erica A. and Belfiore, Francesco and Benincasa, Samantha M. and Bešlić, Ivana and Bigiel, Frank and Bolatto, Alberto D. and den Brok, Jakob S. and Cao, Yixian and Chandar, Rupali and Chastenet, Jérémy and Chiang 江, I-Da 宜 達 and Congiu, Enrico and Dale, Daniel A. and Deger, Sinan and Eibensteiner, Cosima and Egorov, Oleg V. and García-Rodríguez, Axel and Glover, Simon C. O. and Grasha, Kathryn and Henshaw, Jonathan D. and Ho, I-Ting and Kepley, Amanda A. and Kim, Jaeyeon and Klessen, Ralf S. and Kreckel, Kathryn and Koch, Eric W. and Kruijssen, J. M. Diederik and Larson, Kirsten L. and Lee, Janice C. and Lopez, Laura A. and Machado, Josh and Mayker, Ness and McElroy, Rebecca and Murphy, Eric J. and Ostriker, Eve C. and Pan, Hsi-An and Pessa, Ismael and Puschnig, Johannes and Razza, Alessandro and Sánchez-Blázquez, Patricia and Santoro, Francesco and Sardone, Amy and Scheuermann, Fabian and Sliwa, Kazimierz and Sormani, Mattia C. and Stuber, Sophia K. and Thilker, David A. and Turner, Jordan A. and Utomo, Dyas and Watkins, Elizabeth J. and Whitmore, Bradley},
   year={2021},
   month=nov, pages={43} }

@ARTICLE{Schlafy2011,
       author = {{Schlafly}, Edward F. and {Finkbeiner}, Douglas P.},
        title = "{Measuring Reddening with Sloan Digital Sky Survey Stellar Spectra and Recalibrating SFD}",
      journal = {\apj},
         year = 2011,
        month = aug,
       volume = {737},
       number = {2},
          eid = {103},
        pages = {103},
          doi = {10.1088/0004-637X/737/2/103},
archivePrefix = {arXiv},
       eprint = {1012.4804},
 primaryClass = {astro-ph.GA},
       adsurl = {https://ui.adsabs.harvard.edu/abs/2011ApJ...737..103S}
}

@ARTICLE{Pettini2004,
       author = {{Pettini}, Max and {Pagel}, Bernard E.~J.},
        title = "{[OIII]/[NII] as an abundance indicator at high redshift}",
      journal = {\mnras},
         year = 2004,
        month = mar,
       volume = {348},
       number = {3},
        pages = {L59-L63},
          doi = {10.1111/j.1365-2966.2004.07591.x},
archivePrefix = {arXiv},
       eprint = {astro-ph/0401128},
 primaryClass = {astro-ph},
       adsurl = {https://ui.adsabs.harvard.edu/abs/2004MNRAS.348L..59P}
}

@ARTICLE{Kobulunicky2004,
       author = {{Kobulnicky}, Henry A. and {Kewley}, Lisa J.},
        title = "{Metallicities of 0.3<z<1.0 Galaxies in the GOODS-North Field}",
      journal = {\apj},
         year = 2004,
        month = dec,
       volume = {617},
       number = {1},
        pages = {240-261},
          doi = {10.1086/425299},
archivePrefix = {arXiv},
       eprint = {astro-ph/0408128},
 primaryClass = {astro-ph},
       adsurl = {https://ui.adsabs.harvard.edu/abs/2004ApJ...617..240K}
}

@ARTICLE{Kewley2019,
       author = {{Kewley}, Lisa J. and {Nicholls}, David C. and {Sutherland}, Ralph S.},
        title = "{Understanding Galaxy Evolution Through Emission Lines}",
      journal = {\araa},
         year = 2019,
        month = aug,
       volume = {57},
        pages = {511-570},
          doi = {10.1146/annurev-astro-081817-051832},
archivePrefix = {arXiv},
       eprint = {1910.09730},
 primaryClass = {astro-ph.GA},
       adsurl = {https://ui.adsabs.harvard.edu/abs/2019ARA&A..57..511K}
}

@ARTICLE{Kewley2008,
       author = {{Kewley}, Lisa J. and {Ellison}, Sara L.},
        title = "{Metallicity Calibrations and the Mass-Metallicity Relation for Star-forming Galaxies}",
      journal = {\apj},
         year = 2008,
        month = jul,
       volume = {681},
       number = {2},
        pages = {1183-1204},
          doi = {10.1086/587500},
archivePrefix = {arXiv},
       eprint = {0801.1849},
 primaryClass = {astro-ph},
       adsurl = {https://ui.adsabs.harvard.edu/abs/2008ApJ...681.1183K}
}

@Article{         NumPy,
 title         = {Array programming with {NumPy}},
 author        = {Charles R. Harris and K. Jarrod Millman and St{\'{e}}fan J.
                 van der Walt and Ralf Gommers and Pauli Virtanen and David
                 Cournapeau and Eric Wieser and Julian Taylor and Sebastian
                 Berg and Nathaniel J. Smith and Robert Kern and Matti Picus
                 and Stephan Hoyer and Marten H. van Kerkwijk and Matthew
                 Brett and Allan Haldane and Jaime Fern{\'{a}}ndez del
                 R{\'{i}}o and Mark Wiebe and Pearu Peterson and Pierre
                 G{\'{e}}rard-Marchant and Kevin Sheppard and Tyler Reddy and
                 Warren Weckesser and Hameer Abbasi and Christoph Gohlke and
                 Travis E. Oliphant},
 year          = {2020},
 month         = sep,
 journal       = {Nature},
 volume        = {585},
 number        = {7825},
 pages         = {357--362},
 doi           = {10.1038/s41586-020-2649-2},
 publisher     = {Springer Science and Business Media {LLC}},
 url           = {https://doi.org/10.1038/s41586-020-2649-2}
}

@ARTICLE{Teimoorinia2021,
       author = {{Teimoorinia}, Hossen and {Jalilkhany}, Mansoureh and {Scudder}, Jillian M. and {Jensen}, Jaclyn and {Ellison}, Sara L.},
        title = "{A reassessment of strong line metallicity conversions in the machine learning era}",
      journal = {\mnras},
         year = 2021,
        month = may,
       volume = {503},
       number = {1},
        pages = {1082-1095},
          doi = {10.1093/mnras/stab466},
archivePrefix = {arXiv},
       eprint = {2102.07058},
 primaryClass = {astro-ph.GA},
       adsurl = {https://ui.adsabs.harvard.edu/abs/2021MNRAS.503.1082T}
}

@ARTICLE{Marino2013,
       author = {{Marino}, R.~A. and {Rosales-Ortega}, F.~F. and {S{\'a}nchez}, S.~F. and {Gil de Paz}, A. and {V{\'\i}lchez}, J. and {Miralles-Caballero}, D. and {Kehrig}, C. and {P{\'e}rez-Montero}, E. and {Stanishev}, V. and {Iglesias-P{\'a}ramo}, J. and {D{\'\i}az}, A.~I. and {Castillo-Morales}, A. and {Kennicutt}, R. and {L{\'o}pez-S{\'a}nchez}, A.~R. and {Galbany}, L. and {Garc{\'\i}a-Benito}, R. and {Mast}, D. and {Mendez-Abreu}, J. and {Monreal-Ibero}, A. and {Husemann}, B. and {Walcher}, C.~J. and {Garc{\'\i}a-Lorenzo}, B. and {Masegosa}, J. and {Del Olmo Orozco}, A. and {Mour{\~a}o}, A.~M. and {Ziegler}, B. and {Moll{\'a}}, M. and {Papaderos}, P. and {S{\'a}nchez-Bl{\'a}zquez}, P. and {Gonz{\'a}lez Delgado}, R.~M. and {Falc{\'o}n-Barroso}, J. and {Roth}, M.~M. and {van de Ven}, G. and {CALIFA Team}},
        title = "{The O3N2 and N2 abundance indicators revisited: improved calibrations based on CALIFA and T$_{e}$-based literature data}",
      journal = {\aap},
         year = 2013,
        month = nov,
       volume = {559},
          eid = {A114},
        pages = {A114},
          doi = {10.1051/0004-6361/201321956},
archivePrefix = {arXiv},
       eprint = {1307.5316},
 primaryClass = {astro-ph.CO},
       adsurl = {https://ui.adsabs.harvard.edu/abs/2013A&A...559A.114M}
}

@ARTICLE{Dopita2016,
       author = {{Dopita}, Michael A. and {Kewley}, Lisa J. and {Sutherland}, Ralph S. and {Nicholls}, David C.},
        title = "{Chemical abundances in high-redshift galaxies: a powerful new emission line diagnostic}",
      journal = {\apss},
         year = 2016,
        month = feb,
       volume = {361},
          eid = {61},
        pages = {61},
          doi = {10.1007/s10509-016-2657-8},
archivePrefix = {arXiv},
       eprint = {1601.01337},
 primaryClass = {astro-ph.GA},
       adsurl = {https://ui.adsabs.harvard.edu/abs/2016Ap&SS.361...61D}
}

@ARTICLE{Delgado2023a,
       author = {{M{\'e}ndez-Delgado}, J. Eduardo and {Esteban}, C{\'e}sar and {Garc{\'\i}a-Rojas}, Jorge and {Kreckel}, Kathryn and {Peimbert}, Manuel},
        title = "{Temperature inhomogeneities cause the abundance discrepancy in H II regions}",
      journal = {\nat},
         year = 2023,
        month = jun,
       volume = {618},
       number = {7964},
        pages = {249-251},
          doi = {10.1038/s41586-023-05956-2},
archivePrefix = {arXiv},
       eprint = {2305.11578},
 primaryClass = {astro-ph.GA},
       adsurl = {https://ui.adsabs.harvard.edu/abs/2023Natur.618..249M}
}

@ARTICLE{M91,
       author = {{McGaugh}, Stacy S.},
        title = "{H II Region Abundances: Model Oxygen Line Ratios}",
      journal = {\apj},
         year = 1991,
        month = oct,
       volume = {380},
        pages = {140},
          doi = {10.1086/170569},
       adsurl = {https://ui.adsabs.harvard.edu/abs/1991ApJ...380..140M}
}

@ARTICLE{P05,
       author = {{Pilyugin}, Leonid S. and {Thuan}, Trinh X.},
        title = "{Oxygen Abundance Determination in H II Regions: The Strong Line Intensities-Abundance Calibration Revisited}",
      journal = {\apj},
         year = 2005,
        month = sep,
       volume = {631},
       number = {1},
        pages = {231-243},
          doi = {10.1086/432408},
       adsurl = {https://ui.adsabs.harvard.edu/abs/2005ApJ...631..231P}
}

@article{D02,
    author = {Denicoló, Glenda and Terlevich, Roberto and Terlevich, Elena},
    title = "{New light on the search for low-metallicity galaxies — I. The N2 calibrator}",
    journal = {Monthly Notices of the Royal Astronomical Society},
    volume = {330},
    number = {1},
    pages = {69-74},
    year = {2002},
    month = {02},
    issn = {0035-8711},
    doi = {10.1046/j.1365-8711.2002.05041.x},
    url = {https://doi.org/10.1046/j.1365-8711.2002.05041.x},
    eprint = {https://academic.oup.com/mnras/article-pdf/330/1/69/18414396/330-1-69.pdf},
}

@ARTICLE{Pilyugin2001,
       author = {{Pilyugin}, L.~S.},
        title = "{Oxygen abundances in dwarf irregular galaxies and the metallicity-luminosity relationship}",
      journal = {\aap},
         year = 2001,
        month = aug,
       volume = {374},
        pages = {412-420},
          doi = {10.1051/0004-6361:20010732},
archivePrefix = {arXiv},
       eprint = {astro-ph/0105360},
 primaryClass = {astro-ph},
       adsurl = {https://ui.adsabs.harvard.edu/abs/2001A&A...374..412P}
}

@ARTICLE{Delgado2023b,
       author = {{M{\'e}ndez-Delgado}, J.~E. and {Esteban}, C. and {Garc{\'\i}a-Rojas}, J. and {Arellano-C{\'o}rdova}, K.~Z. and {Kreckel}, K. and {G{\'o}mez-Llanos}, V. and {Egorov}, O.~V. and {Peimbert}, M. and {Orte-Garc{\'\i}a}, M.},
        title = "{Density biases and temperature relations for DESIRED H II regions}",
      journal = {\mnras},
         year = 2023,
        month = aug,
       volume = {523},
       number = {2},
        pages = {2952-2973},
          doi = {10.1093/mnras/stad1569},
archivePrefix = {arXiv},
       eprint = {2305.13136},
 primaryClass = {astro-ph.GA},
       adsurl = {https://ui.adsabs.harvard.edu/abs/2023MNRAS.523.2952M}
}

@ARTICLE{Kreckel2019,
       author = {{Kreckel}, K. and {Ho}, I. -T. and {Blanc}, G.~A. and {Groves}, B. and {Santoro}, F. and {Schinnerer}, E. and {Bigiel}, F. and {Chevance}, M. and {Congiu}, E. and {Emsellem}, E. and {Faesi}, C. and {Glover}, S.~C.~O. and {Grasha}, K. and {Kruijssen}, J.~M.~D. and {Lang}, P. and {Leroy}, A.~K. and {Meidt}, S.~E. and {McElroy}, R. and {Pety}, J. and {Rosolowsky}, E. and {Saito}, T. and {Sandstrom}, K. and {Sanchez-Blazquez}, P. and {Schruba}, A.},
        title = "{Mapping Metallicity Variations across Nearby Galaxy Disks}",
      journal = {\apj},
         year = 2019,
        month = dec,
       volume = {887},
       number = {1},
          eid = {80},
        pages = {80},
          doi = {10.3847/1538-4357/ab5115},
archivePrefix = {arXiv},
       eprint = {1910.07190},
 primaryClass = {astro-ph.GA},
       adsurl = {https://ui.adsabs.harvard.edu/abs/2019ApJ...887...80K}
}

@ARTICLE{Pilyugin2012,
       author = {{Pilyugin}, L.~S. and {Grebel}, E.~K. and {Mattsson}, L.},
        title = "{'Counterpart' method for abundance determinations in H II regions}",
      journal = {\mnras},
         year = 2012,
        month = aug,
       volume = {424},
       number = {3},
        pages = {2316-2329},
          doi = {10.1111/j.1365-2966.2012.21398.x},
archivePrefix = {arXiv},
       eprint = {1205.5716},
 primaryClass = {astro-ph.CO},
       adsurl = {https://ui.adsabs.harvard.edu/abs/2012MNRAS.424.2316P}
}

@ARTICLE{Diaz2000,
       author = {{D{\'\i}az}, Angeles I. and {P{\'e}rez-Montero}, Enrique},
        title = "{An empirical calibration of nebular abundances based on the sulphur emission lines}",
      journal = {\mnras},
         year = 2000,
        month = feb,
       volume = {312},
       number = {1},
        pages = {130-138},
          doi = {10.1046/j.1365-8711.2000.03117.x},
archivePrefix = {arXiv},
       eprint = {astro-ph/9909492},
 primaryClass = {astro-ph},
       adsurl = {https://ui.adsabs.harvard.edu/abs/2000MNRAS.312..130D}
}

@article{Vilchez1988,
    author = {Vilchez, J. M. and Pagel, B. E. J.},
    title = "{On the determination of temperatures of ionizing stars in H II regions}",
    journal = {Monthly Notices of the Royal Astronomical Society},
    volume = {231},
    number = {2},
    pages = {257-267},
    year = {1988},
    month = {03},
    issn = {0035-8711},
    doi = {10.1093/mnras/231.2.257},
    url = {https://doi.org/10.1093/mnras/231.2.257},
}

@misc{NIST_elines,
  author       = {A. Kramida and Yu. Ralchenko and J. Reader and NIST ASD Team},
  title        = {NIST Atomic Spectra Database (version 5.11)},
  year         = {2023},
  howpublished = {\url{https://physics.nist.gov/asd}},
  note         = {National Institute of Standards and Technology, Gaithersburg, MD. DOI: \url{https://doi.org/10.18434/T4W30F}},
  urldate      = {2024-07-19}
}

@ARTICLE{Zurita2021,
       author = {{Zurita}, A. and {Florido}, E. and {Bresolin}, F. and {P{\'e}rez-Montero}, E. and {P{\'e}rez}, I.},
        title = "{Bar effect on gas-phase abundance gradients. I. Data sample and chemical abundances}",
      journal = {\mnras},
         year = 2021,
        month = jan,
       volume = {500},
       number = {2},
        pages = {2359-2379},
          doi = {10.1093/mnras/staa2246},
archivePrefix = {arXiv},
       eprint = {2007.12289},
 primaryClass = {astro-ph.GA},
       adsurl = {https://ui.adsabs.harvard.edu/abs/2021MNRAS.500.2359Z}
}

@ARTICLE{Veilleux_1987,
       author = {{Veilleux}, Sylvain and {Osterbrock}, Donald E.},
        title = "{Spectral Classification of Emission-Line Galaxies}",
      journal = {\apjs},
         year = 1987,
        month = feb,
       volume = {63},
        pages = {295},
          doi = {10.1086/191166},
       adsurl = {https://ui.adsabs.harvard.edu/abs/1987ApJS...63..295V}
}

@ARTICLE{Moustakas2010,
       author = {{Moustakas}, John and {Kennicutt}, Jr., Robert C. and {Tremonti}, Christy A. and {Dale}, Daniel A. and {Smith}, John-David T. and {Calzetti}, Daniela},
        title = "{Optical Spectroscopy and Nebular Oxygen Abundances of the Spitzer/SINGS Galaxies}",
      journal = {\apjs},
         year = 2010,
        month = oct,
       volume = {190},
       number = {2},
        pages = {233-266},
          doi = {10.1088/0067-0049/190/2/233},
archivePrefix = {arXiv},
       eprint = {1007.4547},
 primaryClass = {astro-ph.CO},
       adsurl = {https://ui.adsabs.harvard.edu/abs/2010ApJS..190..233M}
}

@ARTICLE{Croxall2013,
       author = {{Croxall}, Kevin V. and {Smith}, J.~D. and {Brandl}, B.~R. and {Groves}, B.~A. and {Kennicutt}, R.~C. and {Kreckel}, K. and {Johnson}, B.~D. and {Pellegrini}, E. and {Sandstrom}, K.~M. and {Walter}, F. and {Armus}, L. and {Beir{\~a}o}, P. and {Calzetti}, D. and {Dale}, D.~A. and {Galametz}, M. and {Hinz}, J.~L. and {Hunt}, L.~K. and {Koda}, J. and {Schinnerer}, E.},
        title = "{Toward a Removal of Temperature Dependencies from Abundance Determinations: NGC 628}",
      journal = {\apj},
         year = 2013,
        month = nov,
       volume = {777},
       number = {2},
          eid = {96},
        pages = {96},
          doi = {10.1088/0004-637X/777/2/96},
archivePrefix = {arXiv},
       eprint = {1309.0817},
 primaryClass = {astro-ph.CO},
       adsurl = {https://ui.adsabs.harvard.edu/abs/2013ApJ...777...96C}
}

@ARTICLE{Peimbert2010,
       author = {{Peimbert}, Antonio and {Peimbert}, Manuel},
        title = "{On the O/H, Mg/H, Si/H, and Fe/H Gas and Dust Abundance Ratios in Galactic and Extragalactic H II Regions}",
      journal = {\apj},
         year = 2010,
        month = nov,
       volume = {724},
       number = {1},
        pages = {791-798},
          doi = {10.1088/0004-637X/724/1/791},
archivePrefix = {arXiv},
       eprint = {1006.0692},
 primaryClass = {astro-ph.GA},
       adsurl = {https://ui.adsabs.harvard.edu/abs/2010ApJ...724..791P}
}

@ARTICLE{Dors2011,
       author = {{Dors}, Jr., O.~L. and {Krabbe}, Angela and {H{\"a}gele}, Guillermo F. and {P{\'e}rez-Montero}, Enrique},
        title = "{Analysing derived metallicities and ionization parameters from model-based determinations in ionized gaseous nebulae}",
      journal = {\mnras},
         year = 2011,
        month = aug,
       volume = {415},
       number = {4},
        pages = {3616-3626},
          doi = {10.1111/j.1365-2966.2011.18978.x},
archivePrefix = {arXiv},
       eprint = {1104.5460},
 primaryClass = {astro-ph.CO},
       adsurl = {https://ui.adsabs.harvard.edu/abs/2011MNRAS.415.3616D}
}

@ARTICLE{Kewley2013,
       author = {{Kewley}, Lisa J. and {Dopita}, Michael A. and {Leitherer}, Claus and {Dav{\'e}}, Romeel and {Yuan}, Tiantian and {Allen}, Mark and {Groves}, Brent and {Sutherland}, Ralph},
        title = "{Theoretical Evolution of Optical Strong Lines across Cosmic Time}",
      journal = {\apj},
         year = 2013,
        month = sep,
       volume = {774},
       number = {2},
          eid = {100},
        pages = {100},
          doi = {10.1088/0004-637X/774/2/100},
archivePrefix = {arXiv},
       eprint = {1307.0508},
 primaryClass = {astro-ph.CO},
       adsurl = {https://ui.adsabs.harvard.edu/abs/2013ApJ...774..100K}
}

@ARTICLE{Kravstov2025,
       author = {{Kravtsov}, T. and {Anderson}, J.~P. and {Kuncarayakti}, H. and {Maeda}, K. and {Mattila}, S.},
        title = "{Discovery of young, oxygen-rich supernova remnants in PHANGS-MUSE galaxies}",
      journal = {\aap},
         year = 2025,
        month = aug,
       volume = {700},
          eid = {A223},
        pages = {A223},
          doi = {10.1051/0004-6361/202349083},
archivePrefix = {arXiv},
       eprint = {2409.06504},
 primaryClass = {astro-ph.HE},
       adsurl = {https://ui.adsabs.harvard.edu/abs/2025A&A...700A.223K}
}

@ARTICLE{Astropy+2013,
  author = {{Astropy Collaboration} and {Robitaille}, Thomas P. and {Tollerud}, Erik J. and others},
  title = "{Astropy: A community Python package for astronomy}",
  journal = {\aap},
  year = 2013,
  volume = {558},
  pages = {A33},
  doi = {10.1051/0004-6361/201322068}
}

@ARTICLE{Astropy+2018,
  author = {{Astropy Collaboration} and {Price-Whelan}, Adrian M. and {Sip{\H o}cz}, B. M. and others},
  title = "{The Astropy Project: Building an inclusive, open-science project and status of the v2.0 core package}",
  journal = {\aj},
  year = 2018,
  volume = {156},
  number = {3},
  pages = {123},
  doi = {10.3847/1538-3881/aabc4f}
}

@ARTICLE{Astropy+2022,
  author = {{Astropy Collaboration} and {Price-Whelan}, Adrian M. and {Lim}, Pey Lian and others},
  title = "{The Astropy Project: Sustaining and growing a community-oriented open-source project and the latest major release (v5.0) of the core package}",
  journal = {\apj},
  year = 2022,
  volume = {935},
  number = {2},
  pages = {167},
  doi = {10.3847/1538-4357/ac7c74}
}

@ARTICLE{Harris+2020,
  author = {{Harris}, Charles R. and {Millman}, K. Jarrod and {van der Walt}, St{\'e}fan J. and others},
  title = "{Array programming with NumPy}",
  journal = {Nature},
  year = 2020,
  volume = {585},
  pages = {357--362},
  doi = {10.1038/s41586-020-2649-2}
}

@ARTICLE{Hunter+2007,
  author = {{Hunter}, John D.},
  title = "{Matplotlib: A 2D graphics environment}",
  journal = {Computing in Science \& Engineering},
  year = 2007,
  volume = {9},
  number = {3},
  pages = {90--95},
  doi = {10.1109/MCSE.2007.55}
}

@ARTICLE{Freedman2001,
       author = {{Freedman}, Wendy L. and {Madore}, Barry F. and {Gibson}, Brad K. and {Ferrarese}, Laura and {Kelson}, Daniel D. and {Sakai}, Shoko and {Mould}, Jeremy R. and {Kennicutt}, Jr., Robert C. and {Ford}, Holland C. and {Graham}, John A. and {Huchra}, John P. and {Hughes}, Shaun M.~G. and {Illingworth}, Garth D. and {Macri}, Lucas M. and {Stetson}, Peter B.},
        title = "{Final Results from the Hubble Space Telescope Key Project to Measure the Hubble Constant}",
      journal = {\apj},
         year = 2001,
        month = may,
       volume = {553},
       number = {1},
        pages = {47-72},
          doi = {10.1086/320638},
archivePrefix = {arXiv},
       eprint = {astro-ph/0012376},
 primaryClass = {astro-ph},
       adsurl = {https://ui.adsabs.harvard.edu/abs/2001ApJ...553...47F}
}

@ARTICLE{Jacobs2009,
       author = {{Jacobs}, Bradley A. and {Rizzi}, Luca and {Tully}, R. Brent and {Shaya}, Edward J. and {Makarov}, Dmitry I. and {Makarova}, Lidia},
        title = "{The Extragalactic Distance Database: Color-Magnitude Diagrams}",
      journal = {\aj},
         year = 2009,
        month = aug,
       volume = {138},
       number = {2},
        pages = {332-337},
          doi = {10.1088/0004-6256/138/2/332},
archivePrefix = {arXiv},
       eprint = {0902.3675},
 primaryClass = {astro-ph.CO},
       adsurl = {https://ui.adsabs.harvard.edu/abs/2009AJ....138..332J}
}

@ARTICLE{Habjan2026b,
       author = {{Habjan}, Eric and {Faesi}, Christopher and {Kreckel}, Kathryn and {M{\'e}ndez-Delgado}, J. Eduardo and {Belfiore}, Francesco and {Vaught}, Ryan J. and {Groves}, Brent and {Scheuermann}, Fabian and {Williams}, Thomas G. and {Klessen}, Ralf S. and {Amiri}, Amirnezam and {Grasha}, Kathryn and {Glover}, Simon},
        title = "{Toward Unbiased Abundance Measurements in Inhomogeneous $\mathrm{H\,II}$ Regions}",
      journal = {arXiv e-prints},
         year = 2026,
        month = jul,
          eid = {arXiv:2607.05295},
        pages = {arXiv:2607.05295},
archivePrefix = {arXiv},
       eprint = {2607.05295},
 primaryClass = {astro-ph.GA},
       adsurl = {https://ui.adsabs.harvard.edu/abs/2026arXiv260705295H}
}
